\documentclass{jhep3}
\usepackage{amsmath, amssymb, amsfonts}
\usepackage{t1enc}
\newtheorem{theorem}{Theorem}
\newcommand{\edth}{\mbox{\symbol{'360}}}

\title{ BMS field theory and holography in asymptotically flat
space-times}

\vspace{1 cm}

\author{Claudio Dappiaggi\\  
Dipartimento di Fisica Nucleare e Teorica,\\

Universit\`{a} degli Studi di Pavia, INFN, Sezione di Pavia, \\
via A. Bassi 6, I-27100 Pavia, Italy\\
E-mail : \email{claudio.dappiaggi@pv.infn.it}}

\preprint{\hepth{0410026}}

\abstract{ We explore the holographic principle in the context of asymptotically flat 
space-times by means of the asymptotic symmetry group of this class of space-times, the so called 
Bondi-Metzner-Sachs (BMS) group. In particular we construct a (free) field theory 
living at future (or past) null infinity invariant under the action of the BMS
group. Eventually we analyse the quantum aspects of this theory and we explore
how to relate the correlation functions in the boundary and in the bulk.}


\begin{document}

\section{Introduction}
Since 't Hooft foundational paper \cite{'tHooft}, the holographic principle 
played a key role in improving our understanding on the nature of gravitational
degrees of freedom in a quantum field theory over a curved background. 
Originally the principle was proposed in order to solve the apparent information
paradox of black holes by means of a theory living on a lower dimensional 
hypersurface (usually the boundary) with respect to bulk space-time where all the 
physical information of the manifold is encoded. 
Moreover, motivated by the Bekenstein entropy formula, the density of data on 
the ``holographic screen'' should not exceed the Planck density which implies 
that there is a high redundancy in the way we usually count degrees of freedom 
in a quantum field theory since if we excite more than $\frac{A}{4}$ degrees of 
freedom, we end up with a black hole.

A way to explicitly realize the holographic principle is based on the 
reconstruction of the bulk starting directly from boundary data explaining how 
they are generated, their dynamics and mainly how they can reproduce classical 
space-time geometry. An example is the AdS/CFT correspondence 
\cite{Aharony} (see \cite{deBoer} for a recent review) where, in the low energy 
limit, a supergravity theory living on $AdS_d\times M^{10-d}$ is a SU(N) 
conformal gauge field theory living on the boundary of AdS. The whole approach is based on the 
assumption of the equivalence of partition sum of gravity and gauge theory once 
asymptotically AdS boundary conditions are imposed on the bulk space-time. Thus 
it seems rather natural to investigate whether we could find a similar 
holographic description once we choose a different class of manifolds and thus 
of boundary conditions. In this paper we will address this problem for 
asymptotically flat space-times, a scenario where the quest for finding a 
suitable holographic description is extremely different from its AdS counterpart.
Just to mention one, a key difference lies in the geometrical aspect of the 
hypersurface where the holographic data are encoded since, while in AdS the 
boundary is a Lorentzian hypersurface, in an asymptotically flat manifold it is 
a null submanifold; this implies a greater difficulty than in AdS if one wishes to 
construct a theory living on the boundary. 

Up to now different approaches have been proposed: in a recent one 
\cite{deBoer2} \cite{Solodukhin}, a Minkowski background is considered and it is 
divided in AdS and dS slices; the idea is to apply separately both AdS/CFT and 
dS/CFT correspondence and then patch together the results. This approach is 
interesting but it is limited up to now only to the flat manifold and it is 
unclear how to extend it to a generic asymptotically flat space-time\footnote{Another 
recent paper \cite{Alvarez} suggests to relate holography in a Ricci flat 
space-time to a Goursat problem i.e. a characteristic problem in a Lorentzian 
setting. }.

In this paper, instead, we continue along the road of \cite{Arcioni}, 
\cite{Arcioni2} where the holographic principle has been explored in asymptotically 
flat space-times by means of the asymptotic symmetry group at null infinity $\Im^\pm$,
namely the Bondi-Metzner-Sachs group (BMS). Since in Penrose intrinsic 
construction of $\Im^\pm$, the BMS is the diffeomorphism group preserving the 
boundary metric, the underlying idea in the previous works was, besides pointing 
out the differences between the flat and the AdS scenario, to study the key 
ingredients of a field theory living on $\Im^\pm$ invariant under a 
diffeomorphism transformation. In particular in \cite{Arcioni}, by means of pure
group theoretical techniques, we have shown the content of the full particle 
spectrum of a BMS field theory (richer than in a Poincar\'e invariant scenario), 
the possible wave functions and the associated equations of motion. Moreover, in 
\cite{Arcioni2}, we have related the (unfaithful) representation of the BMS group
with the infrared sectors of a pure gravitational system at null infinity and we 
have shown that it is possible to construct the dynamic of 
the BMS free particles namely their Hamiltonians; the surprising result is 
the existence of a one to one correspondence between the (covariant) phase space
of BMS and Poincar\'e massive SU(2) particle and between BMS $SO(2)$ and
Poincar\'e $E(2)$ particles. This implies that the boundary theory already at a
classical level encodes all the information from the bulk (Poincar\'e invariant)
theory; nonetheless this is not sufficient to claim about an holographic 
correspondence since the latter is fully manifest at a quantum level and we 
still lack a proper understanding of the quantum BMS field theory not to mention 
a way to reconstruct bulk data (correlation functions in particular) from boundary ones.\\
In this paper, where we adopt the language from the previous works, we address 
this problem starting from the free BMS fields; in particular we construct the 
BMS actions for all the possible wave functions and we study through 
path-integral techniques both the partition function and the 2-point correlation 
function for the leading example of a free massive (and massless) scalar field. 
Let us emphasize that we have chosen the case of a spinless particle only for sake of simplicity but, as 
we will clarify in the paper, the result could be easily obtained for any other 
field. Another natural question concerns the role of interactions both in the 
bulk and in the boundary; we will not directly address this problem in this 
paper since we believe that, before understanding the nature of a candidate 
holographic correspondence for free fields, the above question is premature. 
Nonetheless we wish to emphasize that the a key step in order to analyse the 
holographic counterpart of a bulk interaction  would be to learn how 
interactions could be implemented in the boundary BMS field theory. At a classical level, 
the analysis performed in \cite{Arcioni2} suggests that we have to draw a distinction: 
if we consider a boundary particle such as a BMS scalar field $\phi$ interacting with a non trivial potential (a canonical example is $V(\phi)=\frac{\lambda}{4!}\phi^4$),
the (covariant) phase space can be constructed as in the free field scenario and
the 1:1 correspondence with the bulk counterpart should hold without
modifications as it can be inferred from the analysis in the next section. Instead,
a completely different scenario arises if we consider gauge interactions since
we still lack a way to couple a BMS invariant field with a gauge potential;
techniques such as the ``minimal substitution'', which have been made precise at
a mathematical level in \cite{Sternberg}, \cite{Guillemin} and \cite{Weinstein},
fail in a BMS setting since they strongly rely on symplectic deformation
techniques proper of finite dimensional manifolds; thus the canonical overall
process of substituting in an Hamiltonian framework the momentum $p^\mu$ with
$p^\mu-eA^\mu$ as in QED, cannot be blindly applied to the infinite dimensional
calculus proper of a BMS field theory unless a procedure similar to the one 
proposed by Weinstein in \cite{Weinstein} is developed within this setting. Thus 
the analysis of the role of bulk gauge fields is premature and, in this paper, 
we will not address further this point leaving it for a future analysis. 

Instead, we will propose an answer to the question of the reconstruction of the 
bulk space-time via data on $\Im^\pm$ by means of the so called null surface formulation 
of general relativity. In this peculiar approach it is possible to reproduce up to a conformal factor the metric 
(satisfying Einstein vacuum field equations) starting from a single function 
which is a solution of the light cone equation. This function depends both on a 
bulk point $x^a$ and on the coordinates over $S^2$ and if we keep $x^a$ fixed it 
is defined on null infinity and thus it is a boundary data. Using this peculiar 
formulation of Einstein theory, we will be able to relate boundary and bulk 
2-point correlation function in Minkowski background and we will draw as well 
some conclusions in a generic background.

\vspace{0.3cm}
\begin{center}

{\it Outline of the paper}

\end{center}
\vspace{0.3cm}

\noindent The organization of the paper is the following: in section 2 we review
the construction both of the kinematical and of the dynamical data for the
classical BMS 
field theory. Since all the analysis is performed in a momentum frame, in 
section 3 we switch to the coordinate frame through a peculiar technique proper 
of infinite dimensional calculus. In section 4 we construct the Lagrangian for 
the BMS field and in particular for the specific case of a massive scalar field; 
the results from this section allow us to study via path-integral the partition 
function in section 5 where we also construct the 2-point correlation function. 
In section 6 we formulate a conjecture to relate the boundary with the bulk data 
through the formalism of null surface formulation of general relativity. 
Moreover for sake of completeness there is an appendix on the purely
mathematical details of infinite dimensional calculus that we use 
throughout the paper and there is a second appendix where we review the main 
idea lying behind the null surface formalism. 
\section{BMS fields}
In this section we review some of the key results of our previous papers on the
subject without entering into unnecessary technical details which are fully described in \cite{Arcioni},
\cite{Arcioni2}.\\
Let us consider an asymptotically flat space-time $M$ (see \cite{Wald2} and \cite{Geroch} for a technical definition and analysis);
in the so called Bondi reference frame $y_B=(u=t-r,r,\theta,\varphi)$, the future
null infinity $\Im^+\sim\mathbb{R}\times S^2$ can be endowed with a degenerate metric 
\begin{equation}\label{metric}
ds^2=0\cdot du^2+d\Omega^2,
\end{equation}
where $d\Omega^2$ is the solid angle element. Let us briefly remark that both
the topological and the metric structure of conformal infinity in an
asymptotically flat space-time are universal i.e. there is no trivial physical process
which can change (at least at first order) the structure of $\Im^\pm$ providing that the manifold under
analysis is flat both at past and future null infinity \cite{Geroch2}.\\
Up to a stereographic projection sending the $S^2$ coordinates $(\theta,\varphi)$
in $(z,\bar{z})$ the diffeomorphism group (preserving the so called "strong conformal structure")
of (\ref{metric}) is the so called BMS group i.e.
\begin{gather}
u\longrightarrow K(z,\bar{z})\left(u+\alpha(z,\bar{z})\right),\label{null}\\
z\longrightarrow\frac{az+b}{cz+d}\;\;\;ad-bc=1\label{Poinc},
\end{gather}
where $\alpha$ is any smooth map from $S^2$ to the real axis and 
\begin{equation}\label{ilK}
K=(1+\mid
z\mid)^{-1}\left((az+b)(\bar{a}\bar{z}+\bar{b})+(cz+d)(\bar{c}\bar{z}+\bar{d})\right)^{-1}.
\end{equation}
From (\ref{null}) and (\ref{Poinc}), it is straightforward to recognise that the composition
law for the BMS group has the structure of the semidirect product:
$$BMS_4=SL(2,\mathbb{C})\ltimes N,$$
where $N$ (usually called the {\it supertranslation} subgroup) is the set of maps from $S^2$ into $\mathbb{R}$ endowed with a
suitable topology (see \cite{Arcioni} and \cite{Mc1} for a detailed discussion
on this point). In particular we will choose from now on the Hilbert topology
(i.e. $N=L^2(S^2)$) even though all the results we will achieve 
apply as well to the nuclear topology i.e. $N=C^\infty(S^2)$ which has been
extensively discussed in \cite{Mc4}. Moreover, let us add that there exists a 
unique four dimensional normal subgroup $T^4\subset L^2(S^2)$, but this does not allow us to extract a unique Poincar\'e subgroup from the BMS; on the opposite the number of non equivalent Poincar\'e subgroups is infinite namely there is one for each element $g\in L^2(S^2)/T^4$ since $g^{-1}(SL(2,\mathbb{C})\ltimes T^4)g=SL(2,\mathbb{C})\ltimes T^4$.\\
Since our ultimate goal is to construct a field theory invariant under the action of the BMS
group, the first step is to study the spectrum of the fields living on the
boundary. This result has been achieved in \cite{Arcioni} using the theory of
representations for the BMS group developed in \cite{Mc1} and \cite{Mc2} and we
will now briefly review it. 

A BMS invariant field can be thought as an {\it induced wave equation} i.e. a
map from the orbits in $L^2(S^2)$ of one of the little groups $L_\chi$ of $BMS_4$:
\begin{equation}\label{indeq}
\phi^j:\mathcal{O}=\frac{SL(2,\mathbb{C})}{L_\chi}\hookrightarrow L^2(S^2)\to\mathcal{H}^j,
\end{equation}
where $\mathcal{H}$ is a suitable finite-dimensional Hilbert space usually
chosen as $\mathbb{C}^j$ where $j$ is the dimension of an irreducible
representation of the little group. Each orbit can be described by means of some
characteristic labels, namely the Poincar\'e mass of a particle and the index of
a unitary representation of the little group. Since it will be a key concepts in the
forthcoming sections, let us briefly remember that the Poincar\'e mass of a BMS
particle is defined through the unique four dimensional Abelian normal subgroup
$T^4$ of $BMS_4$. Let us consider any point $\alpha\in L^2(S^2)$ and its
associated character i.e. a map $\chi:L^2(S^2)\to U(1)$ which assigns to
$\alpha$ the phase $\chi(\alpha)=e^{if(\alpha)}$.
Through the Riesz-Fischer theorem, we can write $f(\alpha)$ as the product in
$L^2(S^2)$ between $\alpha$ and a unique element $p(\theta,\varphi)$. In analogy
with a Poincar\'e invariant theory in a finite dimensional manifold $M^d$ where the character associated
to each point $x^\mu\in M^d$ is $\chi(x^\mu)=e^{i p_\mu\cdot x^\mu}$,
we can recognise $p(\theta,\varphi)$ as the (super)momentum associated to the point
$\alpha\in L^2(S^2)$. Moreover if we expand $p(\theta,\varphi)$ in spherical harmonics 
we can easily recognise in the first four harmonics the contribution from the normal
subgroup $T^4$. In detail, let us introduce the following projection
\begin{gather}
\pi:L^2(S^2)\to T^4\notag\\
\pi(p(\theta,\varphi))=\sum\limits_{l=0}^1\sum\limits_{m=-l}^l
p_{lm}Y_{lm}(\theta,\varphi)\longleftrightarrow
p_\mu=(p_{00},p_{1,-1},p_{10},p_{11}).
\end{gather}
which defines the Poincar\'e mass 
$$m^2=\pi(p(\theta,\varphi))\cdot\pi(p(\theta,\varphi))=\eta^{\mu\nu}p_\mu p_\nu.$$
Let us now concentrate on the list of BMS little groups  
($R_2=\left\{I,-I\right\}$):

\begin{center}
\begin{tabular}{|c|c|c|} 
\hline
Little group & orbit invariant & representation label \\
\hline
SU(2) & $p^2=m^2$, sgn($p_0$) & discrete
spin $j$ (dim=2j+1)\\
\hline
$\Gamma\sim SO(2)$ & $p^2=m^2$, sgn($p_0$)& discrete spin $s$\\
\hline
$\Gamma$ & $p^2=0$, sgn($p_0$)& discrete spin $s$\\
\hline
$\Gamma$ & $p^2=-m^2$, & discrete spin $s$\\
\hline
$\Theta\sim SO(2)R_2$ & $p^2=m^2$, sgn($p_0$) & discrete spin $s$\\
\hline
discrete groups & $p^2>=<0$ & finite dim. rep.\\ 
\hline
\end{tabular}
\end{center}
This list has to be compared with the corresponding one for a Poincar\'e
invariant theory:

\begin{center}
\begin{tabular}{|c|c|c|} 
\hline
Little group & orbit invariant & representation label \\
\hline
$SU(2)$ & $p^2=m^2$, sgn($p_0$) & spin $j$ \\
\hline
$SU(1,1)$ & $p^2=-m^2$, & discrete spin $j^\prime$\\
\hline
$E(2)$ & $p^2=0$, sgn($p_0$) & $\infty$-dimensional,\\
\hline
$E(2)$& $p^2=0$, sgn($p_0$) & 1-dimensional $\lambda$. \\ 
\hline
\end{tabular}
\end{center}
At first glance we can recognise several differences between the BMS particle
spectrum and the Poincar\'e spectrum; although in both cases there is an $SU(2)$
invariant massive particle, an $E(2)$ massless
particle and an unphysical $SU(1,1)$ particle characterise a Poincar\'e
invariant theory. On the opposite the BMS
spectrum shows a plethora of fields related to discrete little groups and
massive fields associated to $SO(2)$ and  $\Theta$ little groups whereas the only
possible massless particle comes from the $SO(2)$ representation. \\
Despite these big differences, we can still recognise that all the Poincar\'e
physical spectrum is fully encoded in the BMS spectrum since, as demonstrated in
\cite{Arcioni2}, there is a one to one correspondence between the dynamic
(technically speaking between the covariant phase spaces i.e. the space of
dynamically possible configurations) of SU(2) BMS and Poincar\'e fields and more
surprisingly between BMS $SO(2)$ and Poincar\'e $E(2)$ massless fields.
Moreover there is also a natural interpretation for the discrete little groups
as instantons of the bulk gravitational field. This idea has been introduced in
\cite{Mc7} (see also \cite{Melas} for recent results) but we will not pursue it
further in this paper. Nonetheless a correct interpretation at a level of bulk
data for the massive $SO(2)$ and $\Theta$ fields is still lacking and it is
under investigation. 

\vspace{0.3cm}

\begin{center}
{\it Equation of motion}
\end{center}

\vspace{0.3cm}

A further result we can achieve studying the BMS fields through the theory of
representation are the equations of motion that characterise the
dynamic of each BMS-particle. This result is in some sense the BMS equivalent of
Wigner approach to the relativistic wave equations for Poincar\'e invariant
fields; the key step is to introduce the so called {\it covariant wave equation}
i.e. a function
\begin{equation}\label{coveq}
\tilde{\phi}^\sigma:L^2(S^2)\to\mathcal{H}^\sigma,
\end{equation}
which transforms under the action of the BMS group through a representation of
its Lorentz subgroup (more precisely $SL(2,\mathbb{C})$). The target space
$\mathcal{H}^\sigma$ is a finite dimensional Hilbert space usually chosen as
$\mathbb{C}^\sigma$ where $\sigma$ is the dimension of the chosen
representation of $SL(2,\mathbb{C})$. In this picture the equation of motion
arise realizing that (\ref{coveq}), albeit covariant, is an highly redundant wave function which
transforms under a non-irreducible representation; thus one has to impose some
constraints on (\ref{coveq}) in order to reduce it to (\ref{indeq}) which
transforms under an irreducible representation of the BMS group. These
constraints are twofold; from one side the support
of (\ref{coveq}) has to be reduced from $L^2(S^2)$ to the orbit $\mathcal{O}$ of the
little group $L_\chi$ whereas, from the other side, one has to act
on the target space since the dimension of
$\mathcal{H}^\sigma$ in (\ref{coveq}) is much bigger than the dimension of
$\mathcal{H}^j$ in (\ref{indeq}) i.e. many components of $\tilde{\phi}^\sigma$
are redundant.\\
It is worth to notice that these operations are all performed in a momentum frame
and that for the BMS fields they have been classified and interpreted in \cite{Arcioni}.
Nonetheless, since in the forthcoming chapter we will always refer for sake of simplicity 
to a guiding example, namely the real scalar $SU(2)$ BMS field, let us derive the
equations of motion in this setting. The induced and covariant wave functions are
respectively:
$$\phi:\frac{SL(2,\mathbb{C})}{SU(2)}\sim\mathbb{R}^3\to\mathbb{R},$$
$$\tilde{\phi}:L^2(S^2)\to\mathbb{R}.$$
As shown in \cite{Mc1}, the ``orbit constraint'' is equivalent to the vanishing
of the pure supertranslational part in the supermomentum i.e.
\begin{equation}\label{prima}
p(\theta,\varphi)-\pi(p(\theta,\varphi))=0
\end{equation}
This equation reduces the dynamical space from $L^2(S^2)$ to $\mathbb{R}^4$.
Furthermore, also the mass equation has to be imposed i.e.
\begin{equation}\label{seconda}
[\pi(p(\theta,\varphi))\cdot\pi(p(\theta,\varphi))-m^2]\phi(p)=0,
\end{equation}
which selects the orbit $\frac{SL(2,\mathbb{C})}{SU(2)}$ embedded in
$\mathbb{R}^4$. The set (\ref{prima}) and (\ref{seconda}) determine the full
dynamic of a BMS scalar field and we will refer to them as the {\it
BMS-Klein-Gordon} equations. No further condition has to be imposed on
the target space since the dimension is the same in the induced and covariant
approach. 

In order to convince the reader that the result is meaningful, let us briefly
repeat the construction for the Poincar\'e scalar field. In this scenario the
induced wave function has the same expression as in the BMS case:
$$\psi:\frac{SL(2,\mathbb{C})}{SU(2)}\sim\mathbb{R}^3\to\mathbb{R},$$
whereas the covariant function is
$$\tilde{\psi}:T^4\to\mathbb{R}.$$
The only constraint that has to be imposed is the mass equation:
$$\left[p^\mu p_\mu - m^2\right]\tilde{\psi}(p)=0,$$
which is exactly the Klein-Gordon equation in the momentum frame.

A key feature of Wigner's approach in order to deduce the equations of motion of a
Poincar\'e (or a BMS) particle is the total absence either of a Lagrangian or of an
Hamiltonian function to minimise in order to describe the dynamic. This aspect
is highly interesting since it shows that the dynamic of a free particle is purely
a group-theoretical result but an Hamiltonian (or equivalently a Lagrangian)
approach is nonetheless essential in order to construct the quantum theory from 
a path-integral point of view. In the BMS scenario this problem has been
partially solved in \cite{Arcioni2} where the free Hamiltonians for each BMS
field has been constructed. Considering always the leading example of the 
scalar field let us review the key aspects and results of the construction. In
particular, let us notice that the key ingredient in an Hamiltonian framework is
the correct identification of a phase space; in the usual approach
to quantum field theory this is achieved slicing the 4-dimensional manifold as
$\Sigma_3\times\mathbb{R}$ and recognising $\mathbb{R}$ as the evolution
direction and $\Sigma_3$ as the Cauchy surface where initial data are defined.
Thus the canonical phase space associated to the dynamical theory of a
covariant field $\phi$ is usually identified as the set 
$\Gamma=\left\{\left(\varphi,\pi\right)\right\}$ where $\varphi=\phi|_{\Sigma_3}$ 
and $\pi=\frac{\delta\mathcal{L}}{\delta\phi}\mid_{\Sigma_3}$. This approach is not pursuable
in the BMS case since we lack the identification both of a Lagrangian and of a
suitable initial Cauchy surface in $L^2(S^2)$. Nonetheless an alternative road to
define an Hamiltonian is to consider the so-called {\it covariant phase space}
which is the set of fields satisfying the equations of motion i.e. it is the space
of all dynamically possible configurations. This space is smaller\footnote{There
is only one exception i.e. the Poincar\'e scalar field where the covariant and 
the canonical phase space are in one to one correspondence.} than the
canonical phase space and we refer to the literature (\cite{Lee} and references therein)
for the reader interested in its main characteristics.\\
For a BMS scalar field the covariant phase space is:
\begin{equation}\label{covphase}
\Gamma|_{cov}=\left\{\phi:\frac{SL(2,\mathbb{C})}{SU(2)}\to\mathbb{R}\;\;[\pi(p)\cdot\pi(p)-m^2]\phi(p)=0\right\}
\end{equation}
 Since the constraints defining the equations of motion for a free field are first
 class, we can identify $\Gamma|_{cov}$ as a linear space which can be
 endowed with a suitable regularity condition requiring either that the fields are smooth
 i.e. of class $C^\infty$ or that they are square-integrable respect to a
 suitable measure. In both cases we end up with a covariant phase space which is
 a Banach or an Hilbert space and, thus, we can apply a theorem from Marsden and
 Chernoff (see \cite{Chernoff} and also \cite{streubel} for an earlier
 application in a BMS scenario) which grants us that, if
 we can endow (\ref{covphase}) with a symplectic structure $\Omega$, any Lie algebra
 generator $\zeta$ of $\frac{SL(2,\mathbb{C})}{SU(2)}$ is globally Hamiltonian with
 energy function
 \begin{equation}\label{Hamiltonian}
 H_\zeta(\phi)=\frac{1}{2}\Omega(\mathcal{L}_\zeta\phi,\phi),
 \end{equation}
 where $\mathcal{L}_\zeta$ is the Lie derivative along the $\zeta$ generator.
 Let us further remark that a choice for the symplectic structure on 
 $\Gamma|_{cov}$ can be performed reminding that $\mathcal{O}\sim\frac{SL(2,\mathbb{C})}{SU(2)}$
 is equivalent to $\mathbb{R}^3$ endowed with the hyperbolic metric. Thus we can
 write the canonical
$$\Omega(\phi,\phi^\prime)=\int\limits_{\mathcal{O}}d\mu(\mathcal{O})\left(\phi\nabla\phi^\prime-\phi^\prime\nabla\phi\right),$$
where $\nabla$ is the covariant derivative.
\section{From momentum frame to the coordinate frame: an infinite dimensional
calculus}
As we have explained in the previous section, the pure group theoretical approach  
has allowed us to write the equations of motion for a BMS field in the momentum
frame. Since the aim is to derive these equations from a variational principle,
the first step is to write them in a coordinate frame. Nonetheless, at first
glance, this task is
rather challenging since the space of variables under consideration is not
finite-dimensional and thus we cannot perform the usual Fourier transform
straightforward. In order to circumvent this problem, we have to realize that
the configuration space of fields can be endowed with some regularity constraint
namely we can require it to be a Banach or an Hilbert space. In particular from
now on we will choose it as:
$$\mathcal{C}=\left\{\phi:N=L^2(S^2)\to\mathbb{C}^\lambda\;\;|\;\;\phi\in
L^2(N)\otimes\mathbb{C}^\lambda\right\}.$$
This requirement allow us to activate several interesting techniques which have
been developed in the study of white noise analysis as an infinite dimensional
calculus. In particular the main results we are interested in are, from the pure
mathematical point of view, an extension to an infinite
dimensional setting of key Hilbert space techniques such as
Fourier transform and distribution theory. Since these results are fundamental 
to this paper, we will review them to a certain extent in the
appendix following \cite{Hida} and \cite{Kuo2}. For the less mathematically
oriented reader, we wish to emphasise that, in the BMS scenario, the role of the
finite dimensional manifold $\mathbb{R}^n$ is played by $L^2(S^2)$ which is a
Riemannian infinite-dimensional manifold endowed with a positive definite metric 
defined through the usual
internal product. Thus, for our purposes, the space of square integrable 
function over $L^2(S^2)$ and the operators acting on it
behave exactly as in $L^2(\mathbb{R}^n)$ except for some peculiar aspects which we will
point out as soon as we need them. 

Thus, let us start considering the BMS equations (\ref{prima}) and (\ref{seconda})
for the scalar field in a momentum frame. 
\begin{theorem}
The support condition $[p(\theta,\varphi)-\pi(p)]\phi(p)=0$ is equivalent to
\begin{equation}\label{terza}
Q_{p-\pi(p)}\phi(p)=0,
\end{equation}
where $\phi$ is a non vanishing function in $\mathcal{C}$ and where
$Q_{p-\pi(p)}$ is the multiplication operator in $L^2(N)$ along the
element\footnote{From now on the dependence on the $S^2$ angular variable is
implicit unless stated.} 
$p-\pi(p)$ in $N=L^2(S^2)$. At the same time, if we introduce the basis in $N$,
$\left\{e_i\right\}=\left\{Y_{00},Y_{1-1},Y_{10},Y_{11},...\right\}$, equation (\ref{terza}) is equivalent to 
$$Q_{e_i}\phi(p)=0\;\;\;i=4,...,\infty$$ 
\end{theorem}
In order to demonstrate the above theorem let us remark that, if we expand in
spherical harmonics the supermomentum
$p(\theta,\varphi)=\sum\limits_{l=0}^\infty\sum\limits_{m=-l}^lp_{lm}Y_{lm}(\theta,\varphi)$,
the support condition implies $p_{lm}=0$ for $l>1$. The action of the
multiplication operator is (see appendix A for the definition)
$$Q_{p-\pi(p)}\phi(p)=[<p,p>_N-<p,\pi(p)>_N]\phi(p)=0,$$
where $<,>_N$ is the canonical internal product on $N$. Using the orthogonality
relations between the spherical harmonics and the definition of the projection
$\pi$, the above formula implies
$$\sum\limits_{l>1}\sum\limits_{m=-l}^l\mid p_{lm}\mid^2=0.$$
Since this is a sum of positive quantities, this statement is equivalent to
$p_{lm}=0$ for any $l>1$ which is exactly the support condition for a BMS
$SU(2)$ field. At the same time the request $p_{lm}\phi(p)=0$ for a non
vanishing function $\phi(p)$ can be written as: 
$$<Y_{lm},p>\phi(p)=Q_{Y_{lm}}\phi(p)=0.\;\;\;l>1$$ 
This implies the statement $Q_{e_i}\phi(p)=0$ with
$i>4$.

Equivalently the mass equation for a BMS scalar field can be written through the
multiplication operators. Let us put (\ref{seconda}) in the form
\begin{equation}\label{masseq}
\left[\eta^{\mu\nu}\pi(p)_\mu\pi(p)_\nu-m^2\right]\phi(p)=0,
\end{equation}
where $\eta^{\mu\nu}$ is the flat Lorentzian metric and $\pi(p)_\mu=<\pi(p),e_\mu>$ i.e.
it is the projection of the $T^4$ part of the supermomentum along the
direction in $L^2(S^2)$ defined by the first four harmonics. Let us stress that the explicit presence of the
metric $\eta$ is needed since the above equation characterise the mass hyperboloid
embedded in the Riemannian space $L^2(S^2)$; thus, there is no way to write the
above relation only using the natural Riemannian internal product on the configuration
space and a second Lorentzian scalar product has to be introduced.
\begin{theorem}
The mass equation (\ref{masseq}) is equivalent to:
\begin{equation}\label{masseq2}
[\eta^{\mu\nu}Q_{e_\mu}Q_{e_\nu}-m^2]\phi(p)=0,
\end{equation}
where $\left\{e_\mu\right\}$ is the set of first four harmonics.
\end{theorem}
In order to demonstrate this theorem, we only need to notice that the following
chain of identities holds:
$$\pi(p)_\mu=<\pi(p),e_\mu>=<p,e_\mu>.$$
Thus the mass equation can be written as:
$$\left[\eta^{\mu\nu}<p,e_\mu><p,e_\nu>-m^2\right]\phi(p)=[\eta^{\mu\nu}Q_{e_\mu}Q_{e_\nu}-m^2]\phi(p)=0,$$
where in the second equality we have applied the definition of multiplication
operator acting on $L^2(N)$ with $N=L^2(S^2)$.

We have now written the equation of motion of the Klein-Gordon BMS field through
the action of suitable multiplication operators acting on the covariant wave
function in the momentum frame. The last step consists on using the properties of the
Fourier transform in infinite dimensions in order to switch to the coordinate
frame. Leaving the mathematical details and definitions to the appendix, let us
only stress that the Fourier transform, although defined in a non intuitive way,
shares the same properties with its finite-dimensional counterpart when acting
on operators. Thus, calling with $\mathcal{F}$ the Fourier transform, the
following identities hold:
$$\mathcal{F}Q_\eta=iD_\eta,$$
$$\mathcal{F}D_\eta=iQ_\eta,$$
where $D_\eta$ is the Gateaux (i.e. directional) derivative along $\eta$. Thus,
using these relations, we can write the BMS Klein-Gordon equation in the
coordinate frame:
\begin{eqnarray}
\left[\eta^{\mu\nu}D_{e_\mu}D_{e_\nu}+m^2\right]\hat{\phi}(x)=0,\label{sword}\\
D_{p-\pi(p)}\hat{\phi}(x)=0\label{breaker},
\end{eqnarray}
where $x\in L^2(S^2)$, $\hat{\phi}$ is the Fourier-transformed covariant wave function and where
the second equation implies the vanishing of the gradient of $\phi$ along the 
pure supertranslational directions. This relation can be decomposed in its
components as:
\begin{equation}\label{component}
D_{e_i}\hat{\phi}(x)=0,\;\;\; i>4.
\end{equation}

Thus, we have achieved our goal to write the equation of motion for a scalar
field in the coordinate frame but a natural question that arises is how general
is this construction and how far we can apply it to all other BMS fields. The
answer is that the above construction is completely general and let us make some
comments and remarks on this issue. \\
First of all let us remember that, given a covariant wave function
$\phi^\lambda:\mathcal{O}\sim \frac{SL(2,\mathbb{C})}{L_\chi}\to\mathbb{C}^\lambda$, the
equations of motion in momentum frame are:
\begin{equation}\label{tutte}
\left\{\begin{array}{c}
p-G\bar{p}=0,\\
\left[\pi(p)\cdot\pi(p)-m^2\right]\phi^\lambda(p)=0,\\
\rho^\lambda(p)\phi^\lambda(p)=\phi^\lambda(p),\\
\end{array}\right.
\end{equation}
where $G\bar{p}$ is the action of the full $SL(2,\mathbb{C})$ group on the fixed
point $\bar{p}$ in $L^2(S^2)$ associated to the little group $L_\chi$ \cite{Mc1}. The term
$\rho^\lambda$ is a matrix reducing the redundant components of the wave
function in $\mathbb{C}^\lambda$ (see \cite{Arcioni} for an explicit characterisation and
construction in the BMS scenario). If we want to write these equations in the
coordinate frame, we can immediately draw some conclusions: the mass equation is
identical for every BMS field (both with vanishing or non vanishing mass), thus 
(\ref{sword}) can be thought as a general equation holding for any little group.
This is not surprising if we think to the Poincar\'e case where the mass
equation is simply the Klein-Gordon equation which holds for every covariant
field. 

The orbit constraint instead is dependent on the chosen little group $L_\chi$. Thus
we can immediately conclude that (\ref{breaker}) holds for any $SU(2)$ BMS
particle. Moreover, if we change the little group, one can immediately notice 
without big technical problems that equation (\ref{breaker}) becomes in a coordinate frame
$$D_{p-G\bar{p}}\hat{\phi}^\lambda(x)=0,$$
which grants us that the gradient of the wave function vanishes outside the orbit
of the classical motion. As a side remark, we can also notice that this equation
can be written in the form
\begin{equation}\label{form}
D_{e_i}\hat{\phi}(x)=0,\;\;\; i>dim\frac{SL(2,\mathbb{C})}{L_\chi}+1,
\end{equation}
providing that we perform a suitable non canonical choice of a base in $L^2(S^2)$.
The last term in (\ref{tutte}) has not been discussed in the Klein-Gordon scenario since it
represents the dynamic of the field on the orbit which, in the scalar case, is
fully encoded in the mass equation. Moreover, this term is non universal and it
encodes the differences between the BMS fields; nonetheless we can perform the
same tricks as for the orbit and the mass term in order to write it in the
coordinate frame. To better clarify this point let us consider the spin
$\frac{1}{2}$ BMS field; if we take into account the first four harmonics and if
we follow in a straightforward manner the details of the calculation in
\cite{Barut}, the dynamical equation on the orbit can be written as:
$$[\gamma^\mu p_\mu-m]\phi(p)=0,$$
where $\gamma^\mu$ are the Dirac matrices and where $p_\mu=<e_\mu,p>$ ($e_\mu$
being one of the first four harmonics). Thus, in the language of operators, this
equation is
$$[\gamma^\mu Q_{e_\mu}-m]\phi(p)=0.$$  
Performing a Fourier transform, this equation becomes
$$[i\gamma^\mu D_{e_\mu}-m]\hat{\phi}(x)=0.$$
Thus, the form of the equation of motion for a BMS particle can be written in the
coordinate frame using infinite dimensional techniques and the expression is
rather similar, at least in the $SU(2)$ case, to the Poincar\'e expression. From a
physical point of view, this is not surprising since the main difference between
the Poincar\'e and the BMS approach comes from the pure
supertranslational part which is at classical level sterile since
(\ref{breaker}) grants us that along these directions the covariant field
is constant. This confirms the result, already achieved in \cite{Arcioni2}, that
the dynamic of Poincar\'e fields is fully encoded in the BMS case and there is a
complete equivalence if we consider the $SU(2)$ fields. Nonetheless let us
anticipate that the pure supertranslational term, although only a constraint at a classical
level, will play a fundamental role at a quantum level when performing
path-integration on the full space of configuration and not only on the
covariant one.
A last remark that applies to the $SU(2)$ BMS fields is related to the nature of
the Poincar\'e subgroups of the full BMS group. As we have previously discussed
the latter has a unique four dimensional subgroup but nonetheless we cannot
single out a unique Poincar\'e subgroup. In particular, as we have shown in the
previous section, the number of different Poincar\'e subgroups is infinite i.e. 
equal to the number of elements in $L^2(S^2)/T^4$. At a level of classical 
dynamics, this feature
completely disappears since the orbit of $SU(2)$ BMS-particles is characterised
by a vanishing pure supertranslation selecting in some sense a unique Poincar\'e
subgroup associated to the identity element in $L^2(S^2)/T^4$. We will comment
further on this issue in section 6 when, addressing the problem of
quantisation, the presence of different Poincar\'e subgroups will play a
leading role also in the $SU(2)$ case.

\section{The BMS Lagrangian}
Bearing in mind the results of the previous section, we are now entitled to
investigate from which Lagrangian these equations of motion arise. Let us consider
the BMS Klein-Gordon field and let us start from (\ref{sword}) and
(\ref{breaker}). As underlined before, (\ref{breaker}) is acting as a constraint on
the wave function and it does not provide any dynamic; thus, the function we are
looking for has to be divided in two parts: the first which takes into account
the mass term and the dynamic of the scalar field and the second which takes
into account the constraints through a Lagrange multiplier. In an infinite
dimensional setting the theory of Lagrange multiplier has been studied in
\cite{Schmid} and, using these results, we can introduce the following Lagrangian:
\begin{equation}\label{Lagrangian}
L[\phi]=\phi(x)\left[\eta^{\mu\nu}D_{e_\mu}D_{e_\nu}\phi(x)-m^2\phi(x)\right]+\sum\limits_{i=4}^\infty\gamma_i(x)D_{e_i}\phi(x),
\end{equation}
which gives us the original equation of motion and where $\gamma_i(x)$ are the
Lagrange multipliers. As a side remark and for sake of completeness, we wish to 
point out that the support of each multiplier is for simplicity the whole 
Hilbert space $L^2(S^2)$; this does not grant that the minimum problem in 
(\ref{Lagrangian}) has a unique solution whereas this would have been avoided if 
$\gamma_i(x)$ were defined on the orbit of the little group. Since this issue 
does not play a key role in our analysis, we will not refer to this question 
anymore leaving to \cite{Schmid} for more details.

Following the definition of the Lagrangian, we can introduce as well the Gaussian
measure $d\mu$ in the Hilbert space $L^2(S^2)$ (see \cite{Kuo} for a definition
and properties) in order to write an action
\begin{gather}
S[\phi]=S_{dyn}+S_{constr},\\
\label{action}
S[\phi]=\int\limits_{L^2(S^2)}d\mu(x)\;\phi(x)\left[\eta^{\mu\nu}D_{e_\mu}D_{e_\nu}\phi(x)-m^2\phi(x)\right]+\sum\limits_{i=4}^\infty\gamma_i(x)D_{e_i}\phi(x)
\end{gather}
Let us emphasise a key difference between (\ref{action}) and its finite 
dimensional Poincar\'e-invariant counterpart namely that, whereas through the 
Stokes theorem the Klein-Gordon action can be written as
$$S(\psi)=\int\limits_{M^4}d^4x\sqrt{\mid
g\mid}\left[\partial^\mu\psi\partial_\mu\psi-m^2\psi^2\right],$$
this cannot be done in the BMS case. As a matter of fact, the dynamical part in 
(\ref{action}) can be written as an internal product over the space of fields:
$$S_{dyn}[\phi]=<\phi,\eta^{\mu\nu}(D_{e_\mu}D_{e_\nu})\phi>-m^2<\phi,\phi>.$$
Using the definition of adjoint operator and the relation
$D^*_\eta=Q_\eta-D_\eta$ (see appendix A for a formal definition), the above expression becomes:
$$S_{dyn}[\phi]=\eta^{\mu\nu}<D^*_{e_\mu}\phi,D_{e_\nu}\phi>-m^2<\phi,\phi>=$$
$$=\eta^{\mu\nu}(-<D_{e_\mu}\phi,D_{e_\nu}\phi>+<Q_{e_\mu}\phi,D_{e_\nu}\phi>)-m^2<\phi,\phi>,$$
which, up to a sign, implies:
$$S[\phi]=\int\limits_{L^2(S^2)}d\mu\left[\eta^{\mu\nu}\left(D_{e_\mu}\phi(x)D_{e_\nu}\phi(x)-<e_\mu,x>\phi(x)D_{e_\nu}\phi(x)+\right)\right.$$
$$+m^2\phi^2(x)+\sum\limits_{i=4}^\infty\gamma_i(x)D_{e_i}\phi(x)].$$
\par

As for the equations of motion, a natural question arising from the study of the 
Klein-Gordon action is how general is this construction and how far it can be 
applied to any other free
field. In order to answer to this question we need first of all to separate the
contributions of the Lagrange multipliers from the pure dynamical term. The action
$$S_{constr}(\phi)=\int\limits_{L^2(S^2)}d\mu(x)[\sum\limits_{i=4}^\infty\gamma_i(x)D_{e_i}\phi(x)]$$
is a general term which is dependent only on the little group and, thus, the above
expression holds for any BMS $SU(2)$ field. Moreover similar expressions can be
written for any little group starting from the equation of motion in the
form of (\ref{form}) whereas the tricky point is the dynamical part of the 
action. The latter changes for each different field we consider and a general 
expression cannot be written. Nonetheless the recipe we used in the scalar field scenario is completely general and it can be adapted to any other field providing that an expression for the equation of motion in local coordinates over the orbit is found.

\begin{center}
{\it Massless particles}
\end{center}

\noindent Let us now consider more in detail the case of massless particles that, in a BMS
setting, are associated to $\Gamma\sim SO(2)$ induced wave functions i.e. if we
take a covariant wave function $\psi^\lambda:L^2(S^2)\to\mathbb{C}^\lambda$
the equations of motion are:
\begin{equation}\label{eqmass}
\left\{\begin{array}{c}
[p-G\bar{p}(\theta)]\psi^\lambda(p)=0,\\
\pi(p)\cdot\pi(p)\psi^\lambda(p)=0,\\
\rho^\lambda(p)\psi^\lambda(p)=\psi^\lambda(p),
\end{array}\right.
\end{equation}
where the fixed point $\bar{p}$ is a function only of the $\theta$-variable and
$\rho^\lambda(\bar{p})$ is a diagonal matrix with all 0 entries except the first
equal to $1$. \\
Since we are interested in an holographic description of bulk data,
we are entitled to restrict our analysis to a specific orbit of the $\Gamma$ 
little group with vanishing supertranslation. As we have shown in \cite{Arcioni},
the possible orbits in this peculiar scenario are characterised by the choice of
the ``energy'' for the massless field i.e. the $p_0$ component of $\pi(p)$ and
of a pure supertranslation function\footnote{In the literature, a pure supertranslation refers to an
element of the group $L^2(S^2)/T^4$ i.e. a real function on $S^2$ depending on
spherical harmonics with $l>1$.} $\alpha(\theta)$. Nonetheless the value
$\alpha=0$ has a peculiar meaning since, in this case, it is possible to show
(see \cite{Arcioni2}) that the covariant phase space of a BMS massless $SO(2)$ 
particle is in one to one correspondence with the covariant phase space of a
Poincar\'e massless $E(2)$ particle; in plain words, this implies that the
classical dynamic of a massless bulk field in a Minkowski background is fully
equivalent to the boundary $SO(2)$ BMS counterpart. Thus, in the peculiar
setting, $\alpha=0$ (\ref{eqmass}) reduces to 
\begin{equation}\label{eqmass2}
\left\{\begin{array}{c}
[p-\pi\bar{p}(\theta)]\psi^\lambda(p)=0,\\
\pi(p)\cdot\pi(p)\psi^\lambda(p)=0,\\
\rho^\lambda(p)\psi^\lambda(p)=\psi^\lambda(p).
\end{array}\right.
\end{equation}
Moreover, if we consider the massless real (or complex) scalar field 
i.e. with vanishing helicity, we deal with a one component field 
($\psi^\lambda:L^2(S^2)\to\mathbb{R}$) and $\rho^\lambda(p)=1$. Thus, the
last equation is identically satisfied and (\ref{eqmass2}) is equal to 
(\ref{prima}) and (\ref{seconda}) with the due exception that $m=0$. This
implies that the analysis performed for the massive case can be extended as well
to the $SO(2)$-massless field and from now on all the
results achieved will apply as well to the massless scenario (i.e. considering
simply the limit $m\to 0$) unless stated otherwise.

\section{Path-integral and BMS correlators}
In the previous section we have explicitly constructed the Lagrangian of a
scalar BMS particle and we have given a recipe to perform the same calculation for
any other field. This result allow us to address a key question from the
holographic point of view: what are the main features of the boundary theory at a
quantum level, what is the explicit expression of the correlation
functions and what is eventually their relation (if it exists) with the bulk
counterpart. Seeking an answer to these questions, we will discuss a path-integral
approach to the BMS free fields always through the leading example of the scalar
particle. Thus, starting from (\ref{action}), we need to evaluate:
$$Z[\phi,\gamma]=\int\limits_{\mathcal{C}}e^{iS[\phi,\gamma]}d[\phi,\gamma],$$
where $\mathcal{C}$ is the set\footnote{We wish to emphasise to the reader that
the set $\mathcal{C}$ is not necessarily equivalent to the set of scalar fields defined in the
previous section in order to construct the Lagrangian} of possible kinematical 
configurations endowed with some regularity condition. Remembering that the
action is split in $S=S_{dyn}+S_{constr}$ and that the fields $\gamma_i(x)$ act
only as Lagrange multiplier, we can use the canonical relation due to Faddeev
\cite{Peskin} in order to eliminate $\gamma_i(x)$:
$$Z[\phi]=\int\limits_{\mathcal{C}}d[\phi]\delta\left(\sum\limits_{i=4}^\infty
D_{e_i}\phi(x)\right)
\exp i\left[\phi(x)\left[\eta^{\mu\nu}D_{e_\mu}D_{e_\nu}\phi(x)-m^2\phi(x)\right]\right].$$
The infinite dimensional delta is reminiscent of those appearing in gauge theory
when performing the Faddeev-Popov trick in order to take into account the gauge
degrees of freedom. Thus, we substitute the delta function with the following
combination:
$$Z[\phi]=C\int\limits_{\mathcal{C}}d[\phi]e^{iS_{dyn}}\sum\limits_{i>4}\int d[\omega]e^{\frac{i}{2\zeta_i}
<\omega_i(x),\omega_i(x)>_N}\delta(D_{e_i}\phi(x)-\omega_i(x)),$$
where $C$ is an irrelevant constant, $<,>_N$ is the internal product in $L^2(S^2)$ and $\zeta_i$ is a non
divergent real number. Thus, performing the integration over the
$\omega$-functions, we end up with
$$Z[\phi]=\int\limits_{\mathcal{C}}d[\phi]e^{iS_{eff}},$$
where 
\begin{equation}\label{effaction}
S_{eff}=S_{dyn}+\int\limits_{L^2(S^2)}d\mu(x)\sum\limits_{i>4}[\frac{1}{2\zeta_i}D_{e_i}\phi(x)D_{e_i}\phi(x)].
\end{equation}
In order to evaluate the path-integral for a free field we have to seek if it is
possible to reduce the action in a form $S(\phi)=<\phi,B\phi>_N$. If we look at
(\ref{effaction}) we see that only the second term has to be changed. In
particular each element in the sum can be written as
$$S^\prime=\frac{1}{2\zeta_i}<D_{e_i}\phi(x),D_{e_i}\phi(x)>_N=\frac{1}{2\zeta_i}<\phi(x),D^*_{e_i}D_{e_i}\phi(x)>_N=$$
$$=\frac{1}{2\zeta_i}<\phi(x),(Q_{e_i}-D_{e_i})D_{e_i}\phi(x)>_N,$$
where we have used the definition of adjoint operator and the relation
$D^*_\eta+D_\eta=Q_\eta$.
Thus the operator $B$ is
\begin{equation}
B=\eta^{\mu\nu}D_{e_\mu}D_{e_\nu}+m^2+\sum\limits_{i>4}\frac{1}{2\zeta_i}(Q_{e_i}-D_{e_i})D_{e_i}.
\end{equation}
In analogy with the results for the Poincar\'e free fields, we can formally 
evaluate $Z[\phi]$ up to a constant as a functional determinant:
$$Z[\phi]=const\cdot [det B]^{-\frac{1}{2}}.$$
Let us stress that the above result already shows how the supertranslations act
in a rather peculiar way at a quantum level. Although at a classical level they
are sterile, the functional determinant for the path-integral is not given only
by the Poincar\'e part but the enforcing of the constraint for a vanishing
supertranslation implies the presence of an additional term which contains
surprisingly the multiplication operator.

We can now address the main question in our investigation: the correlation
functions. The first object we need to calculate is the two point function i.e.
$$D(x_1-x_2)=\langle T[\phi(x_1)\phi(x_2)]\rangle=\lim_{T\to\infty}\frac{\int
d[\phi]\phi(x_1)\phi(x_2)\exp\left[i\int\limits_{-T}^Td\mu L\right]}{\int
d[\phi]\exp\left[i\int\limits_{-T}^Td\mu L\right]}.$$
In the BMS scenario the time parameter is not natural to choose as in a
finite-dimensional theory but nonetheless, remembering the
construction of the  Hamiltonian functions, we can use the evolution parameter on the orbit as a natural
candidate. \\
Explicitly, the propagator (or the 2-point correlator) can be calculated as the
inverse of the $B$ operator \cite{Peskin} i.e. in a coordinate frame
$$BD(x_1-x_2)=i\delta(x_1-x_2),$$
where $x_1,x_2$ are two points in $L^2(S^2)$ and where the right hand side is, up
to the pure imaginary constant, the infinite-dimensional delta \cite{Kuo2}.
Performing the Fourier transform, we end up with
\begin{equation}
\left[\eta^{\mu\nu}Q_{e_\mu}Q_{e_\nu}-m^2+\sum\limits_{i>4}\frac{1}{2\zeta_i}(D_{e_i}-Q_{e_i})Q_{e_i}\right]\hat{D}(k)=i.
\end{equation}
Choosing the value of $\zeta_i$ in such a way that
$\sum\limits_{i>4}\frac{1}{2\zeta_i}=0$, the above equation becomes
\begin{equation}\label{propagator}
\left[\eta^{\mu\nu}k_\mu
k_\nu-m^2+\sum\limits_{i>4}\frac{1}{2\zeta_i}[-k^2_i+k_iD_{e_i}]\right]\hat{D}(k)=i
\end{equation}
Eventually we have a formal expression for the 2-point correlation function in
the momentum frame and surprisingly we end up with a first-order differential
equation. Let us make some comments:
\begin{itemize}
\item The (differential) equation for the propagator can be divided in two
parts: the first is identical to the Poincar\'e contribution to the usual 2-point function although the coefficients $k_\mu$ are not a priori related to momenta in the bulk
and the second comes from the constraints i.e. it is a pure supertranslational
effect. Thus we can claim that the result is rather general since
$\sum\limits_{i>4}\frac{1}{2\zeta_i}[-k^2_i+k_iD_{e_i}]$
will appear in all the $SU(2)$ BMS fields and the $\eta^{\mu\nu}k_\mu k_\nu-m^2$
will be substituted with the Poincar\'e counterpart for any other $SU(2)$ field.
\item since (\ref{propagator}) is a differential equation, we need to assign a
suitable initial condition. At a geometric level it is rather interesting to
notice that the initial surface is isomorphic to $T^4$ whereas the pure 
supertranslational directions (i.e. $L^2(S^2)/T^4$) act as the evolution directions. Thus the
initial condition has to be assigned on a four-dimensional surface and the most
natural one is to impose that the propagator has a Poincar\'e-type form for a
fixed value of the supertranslation i.e.
$$\hat{D}(k)=\frac{i}{\eta^{\mu\nu}k_\mu k_\nu-m^2},\;\;
k_i=\bar{k}_i\;\;D_{e_i}D\left(k\right)|_{k_i=\bar{k}_i}=\bar{k}_i
D(k)|_{k_i=\bar{k}_i}.$$
\end{itemize}

At the end of our calculation we have the following result: the two
point function of a BMS field obeys in the momentum frame a differential
equation and the pure supertranslational directions, sterile at a classical
level, are the main responsible for this behavior. Thus, solving
(\ref{propagator}), gives us, roughly speaking, the evolution of the propagator from
one $T^4$ to another or, remembering that there are as many equivalent
Poincar\'e fields as pure supertranslations, from one specific Poincar\'e group
to a second one. This arises several interesting question; at an holographic
level clearly it is imperative to understand how to reconstruct the bulk
information form the boundary theory. Let us stress that the role of $\Im^\pm$
is universal thus the theory living on the boundary is the same for any
asymptotically flat space-time and the BMS field theory cannot distinguish a
priori from one specific metric to another. We will propose a possible solution to these questions
in the forthcoming section. Moreover several interesting analogies arise from
the BMS field theory (at least for $SU(2)$ fields) compared to Ashtekar's
program of asymptotic quantization \cite{asymptoticquantization}; in particular the main variables Ashtekar
proposes to be quantised are the so called $Q_{ab}$-variables which are the
integrals along the degenerate direction on $\Im$ of the news tensor:
$$Q_{ab}(\theta,\varphi)=\int\limits_\Im d^3x N_{ab}(u,\theta,\varphi).$$
The interesting aspect of this approach is that each $Q_{ab}$ carries the
information of a different infrared sector of the gravitational field and,
naively speaking, these are the functions bringing from one Poincar\'e subgroup to
another i.e. they are pure supertranslations. This is exactly the behavior
which appears in our propagator i.e. it is a differential equation bringing us
from one $T^4$ (i.e. a Poincar\'e subgroup) to another and, thus, we could think
to each $T^4$ in (\ref{propagator}) as labeled by a different value of $Q_{ab}$
opening as well the chance to study Ashtekar variable in terms of operators on an
infinite dimensional Hilbert space namely $L^2(L^2(S^2))$. Further relations
between IR sectors of a pure radiative gravitational system and the BMS group
are currently under investigation \cite{workinprogress}
\section{From boundary to bulk: a proposal}
In order to understand if there is a possible holographic relation between a BMS
field theory and the bulk counterpart, the main task is to relate the
correlation functions constructed in the previous section with those living in
the bulk. 

Let us stress that the contents of this section are to a certain degree speculative
and let us start with some remarks: as mentioned before, the BMS group is the
diffeomorphism group for $\Im^+$ which is a universal structure i.e. any
asymptotically flat space-time has the same boundary-structure. This
property extends as well at a level of field theory and correlators, such as
(\ref{propagator}), do not distinguish between two different bulk
manifolds. Nonetheless the BMS-invariance of the boundary theory is a reasonable
request since it is difficult to conceive a quantum field theory that is
not invariant under a diffeomorphism transformation. Thus the path
followed until now is essentially a natural one, but our construction faces some
difficulties: if we look at (\ref{propagator}), we can 
recognise that, even though we could exactly solve the infinite-dimensional 
differential equation, we would face the following questions:
\begin{itemize}
\item The coefficients appearing in the propagator (i.e. $k_i=<e_i,k>$) are 
orthoprojections along spherical harmonics and they have no a priori direct 
natural relation with coordinates of a bulk reference frame. Nonetheless, if we 
remember that in a subclass of asymptotically flat space-times such as Minkowski, 
the BMS group can univocally be reduced to a Poincar\'e group, (\ref{propagator})
with $k_i=0$ (vanishing supertranslation) is formally identical to the scalar 
field propagator in flat background thus suggesting a deeper relation between 
(\ref{propagator}) and the bulk counterpart,
\item BMS fields are defined as maps from $L^2(S^2)$ in
a suitable target vector space $V$. This definition has been a natural choice
since we have followed Wigner approach to construct the particle spectrum,
the equations of motion and the (free field) Hamiltonians from a pure group theoretical point of view. In the Poincar\'e scenario the corresponding choice is \cite{Asorey} 
$$\phi^\lambda:T^4\sim\mathbb{R}^4\to V.$$
This definition is globally correct only in a Minkowski space
where the underlying variety is $\mathbb{R}^4$ endowed with the flat metric
$\eta^{ij}$. In a general framework, the manifold $M^4$ (endowed with a generic Lorentzian
metric $g^{ij}$), where the fields live, is modeled on $\mathbb{R}^4$ and thus
the above definition fits only locally. The same problem applies for the
BMS case where $L^2(S^2)$ in the natural counterpart of $T^4$ but there is no
counterpart for $M^4$ i.e. we lack a natural choice for a global manifold
modeled on $L^2(S^2)$ acting as a configuration space. 
\item in the construction of (\ref{propagator}) there is no notion of the bulk
metric $g^{ij}$. From one side this is a natural consequence of the universality of
$\Im^\pm$ and of the BMS group, whereas, from the other side, from the holographic
point view it is imperative to retrieve information on a non trivial bulk
metric. Let us emphasise that this is a tricky point since we should also take
into account that the presence of $\eta^{\mu\nu}$ in (\ref{propagator}) is a
natural choice for a theory living on the boundary of an asymptotically flat
space-time where the metric is indeed flat.
\end{itemize}

We will now propose a possible solution for the above 
mentioned problems through techniques and results developed in the null 
surface formulation of general relativity.
Since we review its main characteristics in appendix B, we strongly suggest
the reader who is not familiar with this formulation of Einstein theory to read
the appendix before. Let us simply remember that, within this approach, the (conformal) information of a  metric in a 4-dimensional manifold $M^4$ can be reconstructed starting from a characteristic function \cite{Kozameh}
\begin{equation}\label{laZ}
u=Z(x^a,z,\bar{z}),
\end{equation}
where $x^a$ is a point lying in $M^4$, $(u,r,z,\bar{z})$ is a Bondi coordinate 
system in the neighbourhood of $\Im$ and (\ref{laZ}) is the solution in the 
variable $u$ of the null cone equation $L(x_a,x_a^\prime)=0$ where $x_a^\prime\in\Im$. 
As explained in detail in appendix B, (\ref{laZ}) encodes all the information of 
the metric in the bulk manifold $M^4$ and it can be used together with a scalar 
function $\Omega:M^4\times S^2\to\mathbb{R}$ as key ingredients in an alternative 
formulation of Einstein's equations. 
Moreover $Z(x^a,z,\bar{z})$ suits for different interpretations: for an 
arbitrary but fixed value of the pair $(z,\bar{z})$, it represents a null 
surface with respect to the metric in $M^4$ whereas, if $x^a$ is kept fixed, it 
represents real a 2-surface lying on $\Im^\pm$. Thus, within this latter 
framework, (\ref{laZ}), also called {\it cut function} $\mathcal{C}_x$, analytically describes the intersection of the light cone with apex $x^a$ with $\Im^+$ and it is a boundary data. Moreover, since these functions are constructed inverting the light cone equation, there exists always a suitable
open set near $\Im^+$ where $Z(x^a,z,\bar{z})$ is a smooth function and, thus, it
lies in $C^\infty(S^2)$ (or $L^2(S^2)$ if we introduce a further integrability
condition) although in general they are globally neither differentiable neither
single-valued. These remarks suggest us to adopt a new point of view in our
analysis of a boundary BMS-invariant field theory choosing the set of cut
functions (parametrically dependent on a bulk point $x_a$) as the  configuration
space\footnote{This choice could be thought as switching from an element of
$T^4$ to a generic point in $M^4$ in a Poincar\'e scenario.}; this 
also implies a solution for the second of the problems outlined at the 
beginning of the section.  

\vspace{0.3cm}

\begin{center}
{\it Behaviour under a BMS transformation}
\end{center}

\vspace{0.3cm}

\noindent The first key step is to understand the behaviour of cut functions 
under a BMS transformation. Although, not within the perspective of an application in holography, this problem has been analysed in \cite{Frittelli6} and mainly in \cite{Frittelli7} where it was provided a description for the construction of null surfaces and their singularities through
the eikonal equation. The starting point is (\ref{laZ}), a section of the null bundle $S^2\times\mathbb{R}$, which satisfies the eikonal equation
$$g^{ab}\partial_a Z\partial_b Z=0,$$
and which, in a more physical language, represents the family of all asymptotic 
plane waves. An important feature of the above equation is that, if a special 
solution, say $Z_0(x^a,z,\bar{z})$, is known, then all the solutions are known 
as well and, in particular, we can construct all the characteristic (level) surfaces
of some $Z$ given by $u=Z=const$ through the equation
\begin{equation}\label{BMS1}
Z(x^a,z,\bar{z})=Z_0(x^a,z,\bar{z})+\alpha(z,\bar{z}),
\end{equation}
together with $\partial [Z-\alpha]=0$ and $\bar{\partial}[Z-\alpha]=0$. In an 
asymptotically flat space-time the special solution can be chosen 
as the cut function $u=Z(x^a,z,\bar{z})$ that we have introduced before and 
(\ref{BMS1}) describes its behaviour under the action of a supertranslation. 
Moreover, in this scenario, we can also introduce important variables:
$$\Lambda=\partial\left\{(1+z\bar{z})\partial Z\right\},\;\;\;R=(1+z\bar{z})^2\partial\bar{\partial}Z,$$
representing respectively the acceleration along $(z,\bar{z})$ constant curves and the extrinsic 
curvature of the light cone cut; they also allow for an explicit determination of caustics through the 
vanishing of the following determinant:
\begin{equation}\label{dist}D=\left|\begin{array}{cc}
\Lambda & R \\
R & \bar{\Lambda}\end{array}	\right|=0\end{equation}
Surprisingly it is more tricky to understand the effect of a Lorentz transformation on a cut
function. This issue has been addressed in \cite{Frittelli8} and to some extent
in \cite{Frittelli} and \cite{Frittelli4} where it was shown that, acting with (\ref{Poinc}) on $Z$, it
transforms as:
$$Z^\prime(x^\prime_a,z^\prime,\bar{z}^\prime)=K\left(Z(x^a,z,\bar{z})\right),$$
where $K$ is given by (\ref{ilK}). If we expand in spherical harmonics and if we
separate the contribution of the first four harmonics, we end up with
$$Z=x^al_a(z,\bar{z})+\sum\limits_{l\geq 2}z_{lm}Y_{lm}(z,\bar{z}),$$
where $l^a$ is the 4-vector $(Y_{00},...,Y_{11})$. Expanding in a similar way
$Z^\prime$ i.e.
$$Z=x^{\prime a}l_a(z^\prime,\bar{z}^\prime)+\sum\limits_{l\geq 2}z^\prime_{lm}Y_{lm}(z^\prime,\bar{z}^\prime),$$
we can see that a Lorentz transformation maps the point $x^a$ to
\begin{equation}\label{active}
x^{\prime a}=\Lambda^a_bx^b+\Lambda_{lm}(\Lambda^a_b)z^{lm}(x^a).
\end{equation}
The interpretation of the role of $x^\prime_a$ is rather tricky since, if we
consider the Lorentz transformation in a passive sense, then $Z^\prime$ is
simply the same characteristic function seen from a new coordinate system and,
thus, it still represents the light cone with apex $x^a$. On the opposite, if we
give an active sense to the Lorentz transformation, i.e. we interpret
(\ref{active}) as a new space-time point, we are promoting $x^\prime_a$ to a new
bulk point and the Lorentz transformation plays the role of a diffeomorphism
of the interior manifold; within this approach we pay the price that the
original cut function is mapped to a different cut which is not in general the
future light cone of the point $x^\prime_a$. Thus this implies that a boundary
Lorentz transformation (and more generally a Poincar\'e transformation) cannot
be promoted to a bulk one and in a geometrical language this implies (see
\cite{Frittelli8}) that, although at $\Im^\pm$ an asymptotically flat space-time
$M^4$ do indeed become flat, there is no unique Minkowski space-time that can be
associated to $M^4$ in a neighbourhood of future (or past) infinity\footnote{In some sense this is related to the property of the BMS group that it does not
exists a unique Poincar\'e subgroup and the origin of this peculiarity comes
exactly from the higher harmonics in the expansion in spherical harmonics of a
supertranslation.}. Thus we can interpret the coordinates $x^a$ as associating
for every fixed Bondi reference frame a flat space-time with the same $\Im$; 
space-time points are labeled by $x^a$ and a Minkowskian metric
$\eta^{ij}$ is introduced in addition to the bulk one $g^{ij}$.   

\vspace{0.3cm}

\begin{center}
{\it Cut functions and holography}
\end{center}

\vspace{0.3cm}

\noindent Since our aim is to consider the cut functions as the fundamental variables for the boundary theory, let us make some remarks on their properties from an holographic perspective:

\begin{itemize}
\item the function $Z(x^a,z,\bar{z})$ is a non local variable \cite{Kozameh}.
This is an important feature since we require that the theory living on $\Im$ satisfies the holographic
entropy bound which cannot be reproduced by an ordinary local quantum field theory.
\item if we consider a subclass of space-times ``sufficiently close to
Minkowski'', namely asymptotically simple 
manifolds where vacuum Einstein equations are imposed and asymptotically flat 
space-times obtained from hyperboloidal initial data\footnote{It has been 
conjectured that this class of manifolds is a subset of asymptotically simple 
space-times although this has not been yet demonstrated.} \cite{Frittelli4}, the 
first four coefficients of the spherical harmonics expansion of (\ref{laZ}) (i.e. $k_i=<e_i,Z>$ with $e_i=Y_{lm}(\theta,\varphi)\;l=0,1$) define a canonical global coordinate system for the whole space-time.
Let us emphasise that, although  in the above remark we restrict the analysis to a 
subset of an asymptotically flat manifolds, nonetheless it represents the 
collection of the most physical relevant manifolds and, thus, we are not losing sensible informations performing the above choice. 
\item as explained more in detail in appendix B, the set of cut functions allows 
for a direct reconstruction of the conformal data of the bulk space-time. 
Thus we can look at (\ref{laZ}) either only as a kinematical data for the 
boundary theory either as a dynamical one satisfying (\ref{eins}, \ref{uno},
\ref{due},\ref{tre}) which are 
equivalent to Einstein vacuum equations.
\end{itemize}
After these remarks, let us now consider 
the space of cut functions $\left\{\mathcal{C}_x\right\}$ (with the bulk point 
$x^a$ acting as a parameter) as our configuration space and let us analyse the 
consequence in a BMS invariant field theory with the following modification:
$$\phi^\lambda:L^2(S^2)\to
V\longrightarrow\phi^\lambda:\left\{\mathcal{C}_x\right\}\to V,$$
where $V$ is a suitable target vector space. Due to the difficulties to 
explicitly construct (\ref{laZ}) for a fixed space-time, we will first refer to 
Minkowski background and, after, we will make some considerations for a general bulk manifold.

\vspace{0.3cm}

\begin{center}
{\it Minkowski space-time}
\end{center}

\vspace{0.3cm}

\noindent Let us start now considering the leading example of an $SU(2)$ scalar field:
$$\phi:\left\{\mathcal{C}_x\right\}\to\mathbb{R},$$
and, for sake of simplicity, we also require that the set of null surfaces
$\left\{C_x\right\}$ is constituted only by functions satisfying a suitable 
regularity condition. This request is not very restrictive since as mentioned 
before, cut functions are always smooth at least in a suitable neighbourhood of 
$\Im$ which implies that we can apply the formalism used in the previous section 
in a straightforward manner. The only subtlety comes from the necessity to 
introduce the equivalent of the supermomentum space in the pure BMS setting; the 
dual space can be constructed as the set of functions $Z^*(k^a,z,\bar{z})$ 
(either smooth or square integrable in the complex variables) depending on the 
4-dimensional bulk momentum $k^a$.\\
Thus, repeating the calculation as in section 5, we end up with a 2-point 
function as in (\ref{propagator}) whose argument now depends on $Z^*_k$ 
(or rather on its projections along the spherical harmonics):
$$\left[\eta^{\mu\nu}k_\mu
k_\nu-m^2+\sum\limits_{i>4}\frac{1}{2\zeta_i}[-k^2_i+k_iD_{e_i}]\right]\hat{D}(Z^*_k)=i,$$
with $k_\alpha=<Z^*(k^a,z,\bar{z}),e_\alpha>$.
As mentioned before, this is a functional differential equation and, as far as 
we know, there is no way to solve it analytically. Nonetheless, in the peculiar 
case of a flat background, the cut-function has been constructed and it has a simple form \cite{Kozameh},\cite{Frittelli7}:
\begin{gather}
Z(x_a,z,\bar{z})=x_al^a(z,\bar{z})\label{lcmin}\\
l^a(z,\bar{z})=\frac{1}{\sqrt{2}(1+z\bar{z})}\left((1+z\bar{z}),
-(z+\bar{z}),i(z-\bar{z}),(1-z\bar{z})\right),\label{ellea}
\end{gather}
where $x^a$ is the canonical Minkowskian global coordinate system. Moreover, in 
order also to understand the simplicity of (\ref{lcmin}), we have to take into 
account that, in a flat manifold, the BMS group can be univocally reduced to the Poincar\'e group which is reflected in the above formula through the absence of a pure supertranslational part. 
Thus, switching to the (super)momentum frame changing the $x_a$-variable with 
$k_a$, we can substitute $Z^*=k_al^a(z,\bar{z})$ in  (\ref{propagator}); since 
each $k_i=0$ and since we can interpret the coefficients $k^a$ of the first four 
harmonics as a Minkowskian coordinate frame, we end up exactly with the 
Poincar\'e propagator:
$$D(k)=\frac{i}{\eta^{\mu\nu}k_\mu k_\nu +m^2}.$$

\vspace{0.3cm}

\begin{center}
{\it General background}
\end{center}

\vspace{0.3cm}

\noindent In a general background the analysis of the boundary theory and in 
particular of (\ref{propagator}) is rather more complicated even after the 
introduction of cut functions. If we consider a four dimensional asymptotically 
flat manifold $(M^4,g^{\mu\nu})$, the first step would be the construction of
possible cut functions in the chosen space-time and this operation can be 
technically difficult. At present, $Z(x^a,z,\bar{z})$ has been successfully 
analysed only in few specific examples namely for the Schwartzschild and the 
Kerr metric \cite{Joshi}, \cite{Joshi2}. In order to better clarify this difficulty and in 
order to leave the paper self-consistent, we briefly review the construction of
the cut function for the Schwartzschild case; in a Bondi reference frame $(u,r,\theta,\varphi)$,
where $u=t-r-2m\log(r-2m)$, the metric is
$$ds^2=(1-\frac{2m}{r})du^2+2dudr-r^2(d\theta^2+\sin^2\theta d\varphi^2),$$
which can be conformally transformed by $\Omega=r^{-1}=l$ in
\begin{equation}\label{Schwarmet}
d\hat{s}^2=\Omega^2ds^2=4(l^2-\sqrt{8}ml^3)du^2-4dudl-(d\theta^2+\sin^2\theta
d\varphi^2).
\end{equation}
In order to simplify the calculation, it is worth to notice that, due to spherical symmetry, 
we can first investigate the light cone equation on the equatorial plane $\theta=\frac{\pi}{2}$ 
and then we can generate all the possible solutions by a rigid rotation. Thus, starting form 
(\ref{Schwarmet}) and switching from $(\theta,\varphi)$ to complex coordinates $(z,\bar{z})$ 
via stereographic projection, the light cone equation is
\begin{gather}
2(l^2-\sqrt{8}ml^3)\dot{u}-\dot{l}=1,\;\;\ddot{u}+2(l-\sqrt{18}ml^2)\dot{u}^2=0,\\
\ddot{\varphi}=0,\;\;\dot{\varphi}=b,\;\;(l^2-\sqrt{8}ml^3)\dot{u}^2-\dot{u}\dot{l}=\frac{b^2}{4},
\end{gather}
where $b$ is a constant. The above equations can be written in the following form:
\begin{gather}\label{eq1}
\dot{u}=\frac{1+\dot{l}}{2(l^2-\sqrt{8}ml^3)},\;\;\dot{l}=\pm(\sqrt{8}mb^2l^3-b^2l^2+1)^{\frac{1}{2}}=\pm\sqrt{A},\\
ds=\pm\frac{dl}{\sqrt{A}},\;\;d\varphi=\pm\frac{b}{\sqrt{A}}dl, \label{eq2}
\end{gather}
Keeping fixed the light cone apex $x_a=(u_0,l_0,z_0,\bar{z}_0)$, we can realize from (\ref{eq1}) that the possible solutions to
the above differential equations are divided into two sets depending on the sign of $\dot{l}$.
If we consider $\dot{l}<0$, we can integrate (\ref{eq1}), (\ref{eq2}) and, after a lengthy calculation, the final
result is
\begin{equation}\label{equ}
u=u_0-\frac{1}{2}\int\limits_{l_0}^0\frac{b^2}{\sqrt{A}}dl^\prime+\frac{1}{2}\int\limits_{l_0}^0\frac{b^2}{\sqrt{A}+1}dl^\prime,
\end{equation}
$$\varphi-\varphi_0=-\int\limits_{l_0}^0\frac{b}{\sqrt{A}}dl^\prime.$$
In order to obtain the cut function, we have to perform the rotation along the $\theta$-direction and 
this can be achieved substituting the latter equation with
\begin{equation}\label{eqz}
\varphi-\varphi_0=arccos(1-l_a(z,\bar{z})l^a(z_0,\bar{z}_0))=-\int\limits_{l_0}^0\frac{bdl^\prime}{\sqrt{A}},
\end{equation}
where $l_a$ is given by (\ref{ellea}). Since we cannot solve explicitly the above elliptic integrals, 
it is impossible to give an explicit expression for $u=Z(x^a,z,\bar{z})$ and the latter
can be written only parametrically via (\ref{equ}) and (\ref{eqz}).

Although a similar solution can be calculated for $\dot{l}>0$, this would not 
improve our understanding of the construction and of the properties of the cut
function in the Schwartzschild background. Instead, in the spirit of finding an 
holographic correspondence in an asymptotically flat space-time, it is important 
to stress that the above procedure confirms the peculiar nature of the flat 
scenario\footnote{Let us notice that, in the limit of a vanishing Schwartzschild 
mass $m\to 0$, (\ref{equ}) and (\ref{eqz}) converge to the Minkowski cut 
function (\ref{lcmin}).} where it is easier to study an holographic 
correspondence due to the presence of an explicit analytic cut function. 
Nonetheless, although in the Schwartzschild case, the boundary data can be 
written only in a parametric form, it is worth to notice that we still deal with 
a cut function without any supertranslational component; thus, despite the 
presence of elliptic integrals, it is still conceivable, in a future research, to 
look for a solution to the propagator differential equation (\ref{propagator}) 
in this background. As a final remark, let us emphasize that the absence of any 
spherical harmonic with $l>1$ should not surprise since it is known that the BMS 
group could be univocally reduced to the Poincar\'e group by means of geometric 
arguments in any stationary space-time and the equations (\ref{equ}) and 
(\ref{eqz}) simply reflect this peculiar property.

A second step, in order to study holography in an asymptotically flat manifold via null surface formulation of general relativity, consists in the study of the differential 
equation for the propagator in order to construct an explicit solution of 
(\ref{propagator}); although, as mentioned before several times, we lack a 
technique to complete this task whenever the BMS group cannot be reduced in a 
unique way to the Poincar\'e group as in Minkowski space,
we can nonetheless draw some interesting conclusions: first of all, with the 
introduction of cut functions, we can interpret the coefficients $k_\mu$ as 
global coordinates in the bulk manifold and, since the pure supertranslations 
act as evolution directions, we can interpret the differential equation as 
connecting different Minkowski space-times. We can also be tempted to
conjecture that the BMS field theory, together with 
the cut functions, allows for the reconstruction of the bulk data in a neighbourhood of future (past) infinity.
A natural question arising immediately is whether it is conceivable to extend 
the reconstruction of bulk data far beyond those points where the metric can be 
approximated to a good degree to $\eta^{\mu\nu}$. Although an answer to this 
question might be premature, we wish nonetheless to conjecture a possible 
solution which consists in the following picture: until now, the set of cut 
functions acted as a pure kinematical data. Nonetheless, as explained in 
appendix B, we can think to $Z(x^a,z,\bar{z})$ as a dynamical data allowing us 
to reconstruct univocally the (conformal) bulk metric and satisfying a certain 
set of equations equivalent to Einstein's vacuum equations (\ref{eins}, 
\ref{uno}, \ref{due}, \ref{tre}) for general relativity. Switching to this point 
of view, one is entitled to substitute in the BMS action the flat metric with a 
generic metric which is no more an independent data, but it is a now dependent on the cut functions i.e.
\begin{gather}\notag
S=\int\limits_{L^2(S^2)}d\mu(Z_x)\phi(Z_x)\left[g^{\mu\nu}(Z_x)D_{e_\mu}D_{e_\nu}\phi(Z_x)-m^2\phi(Z_x)\right]+\\
+\sum\limits_{i=4}^\infty\gamma_i(Z_x)D_{e_i}\phi(Z_x).\label{actcurv}
\end{gather}
Let us emphasise that this modification does not change any result at a classical level with the due exception of the mass relation:
$$m^2=\eta^{\mu\nu}p_\mu p_\nu\to m^2=g^{\mu\nu}(Z_x)p_\mu p_\nu.$$
Thus we believe that it could be interesting to study if, (\ref{actcurv}) 
together with the equations for the cut functions (\ref{eins}, \ref{uno}, 
\ref{due}, \ref{tre}), allows for a complete holographic reconstruction of bulk 
data.

\section{Conclusions}
In this paper we have studied some aspects of a quantum BMS field theory and
the relations with
an holographic correspondence in asymptotically flat space-times. In particular, through the leading example of a massive BMS scalar field we have succeeded in constructing an action for this particle and we have studied its partition and 2-point correlation function. Surprisingly the latter turned to be quite different from its counterpart in an ordinary Poincar\'e invariant quantum field theory since it is a differential equation. Far from being able to analytically solve it, we have nonetheless draw 
some conclusions namely the pure supertranslational components of the 2-point function, which are sterile at a classical level, act as evolution directions. Thus we have been able to interpret (\ref{propagator}) as an equation connecting the value of the propagator from one Poincar\'e subgroup of the BMS to another; from one side this is not surprisingly since there is not a unique way to construct a Minkowski background in a neighbourhood of $\Im$, namely one for each element of $L^2(S^2)/T^4$, whereas from the other side it seems interesting to analyse such result from the perspective of Ashtekar asymptotic quantisation program \cite{asymptoticquantization} where a key ingredient is played by infrared sectors each of them labeled by a pure supertranslational element. Thus it seems worth on its own for a future work to study BMS field theory and its characteristics despite its relation with holography. On this side instead, the construction of BMS actions and correlation functions has emphasised that BMS invariance plays a necessary but not a sufficient 
role to completely understand a possible bulk to boundary correspondence. In 
particular the role of supertranslation $L^2(S^2)$ as main kinematical data 
seems lacking insights on how to fully understand the boundary theory in the 
same way as $T^4$ does not allow to plainly understand a Poincar\'e invariant 
theory in an (asymptotically) flat background. 

Thus we have introduced a new ingredient in our analysis: cut functions i.e.
maps $Z:M^4\times S^2\to\mathbb{R}$. As explained in the previous section and in
appendix B, in the null surface formulation of gravity they allow for a full
reconstruction of the metric and of Einstein's equations and, more important from
our point of view, they can be interpreted as boundary data. Using these
functions as the main variables and keeping in mind that the first four
coefficients of the spherical harmonics expansion of (\ref{laZ}) define a global
pseudo Minkowskian coordinate system, we have been able to relate the BMS
correlation function with its bulk counterpart in a flat background.
 
In order to clarify the overall picture and, since in this paper we have always 
referred to a boundary scalar field, we wish 
to briefly switch the perspective commenting on how the reconstruction of a bulk 
scalar field works in the language of null surface formulation. If we consider a
Minkowski background, the classical dynamic of a free bulk scalar fields is completely
equivalent to its BMS counterpart. As we have shown in the previous sections,
the pure supertranslational components are sterile and the BMS-Klein-Gordon equation
is defined on the mass hyperboloid as in a Poincar\'e invariant
theory. The big difference instead lies at a quantum level; since the quantum
BMS field theory does not distinguish between different asymptotically flat
space-times, the data of a bulk scalar field can be reconstructed only if we
add to the boundary BMS scalar field a specific choice of a cut
function $Z(x^a,z,\bar{z})$. As we have shown in section 6, the latter data allows for a selection
in (\ref{propagator}) of a specific solution of the differential equation and
the correct formula for the 2-point correlation function for a free scalar field
living in a flat manifold is retrieved.
In a more generic setting, instead, the difficulties 
we face in order to solve the differential equation has not
allowed us to draw a similar conclusion although it seems rather natural to
conjecture that a similar procedure holds for every asymptotically flat space-time
and for any free field.
To be more precise, we need to emphasise that Minkowski is in a lot of ways a
peculiar example since we can reconstruct the whole bulk with a single
square-integrable cut function. On the opposite in a generic manifold, the 
function $Z(x^a,z,\bar{z})$ is neither single valued neither smooth except for a 
suitable neighbourhood of $\Im^+$. Thus, in this scenario, we do not expect to 
reconstruct the whole bulk/boundary correspondence from a single cut function; 
furthermore, in future analysis, we will need to overcome two possible 
difficulties: the first is rather technical and it is related to the fact that 
light cones develop caustics i.e. points where the description of the space-time 
through the eikonal function fails and these points can be characterised by the divergence 
of (\ref{dist}). A second problem comes from the flat metric which is intrinsic 
in the BMS action; this is a natural ingredient for a theory living on $\Im^\pm$ 
where the metric is indeed $\eta^{\mu\nu}$ but, it seems to us that, if we wish to 
reconstruct the data of fields living on a curved bulk background, we lack the 
information from the metric itself. A possible way to overcome this problem has 
been suggested at the end of the previous section substituting $\eta^{\mu\nu}$ 
with a generic metric $g^{\mu\nu}(Z)$ which is dependent on the cut function 
itself. This trick would in principle allow us to solve the above mentioned 
problem but calculations for the propagator appear now rather more difficult. 

As a final remark, we wish to comment on the construction of an S-matrix within
the setting we have considered. As it has been pointed out in \cite{Arcioni},
the overall picture, which emerges from a BMS approach to holography, is rather
similar to 't Hooft analysis in the context of black holes (see \cite{'tHooft2}
and \cite{Arcioni4} for a recent review and developments).
Without entering into details, in this latter scenario, an S-matrix ansatz is 
assumed and the description is given in a first quantized set up; the
holographic fields live on the future and on the past horizon and they depend
only upon angular coordinates. If we consider two such fields, namely
$u(\theta,\varphi)$ and $v(\theta,\varphi)$, the high degree of non locality of
this approach is fully encoded in the operator algebra
$$[u(\theta,\varphi),v(\theta^\prime,\varphi^\prime)]=if(\theta-\theta^\prime,\varphi
-\varphi^\prime),$$
where $f$ is the Green function of the Laplacian operator acting on the angular
horizon coordinates. If we take into account that non locality is a
fundamental request of an holographic theory and if we consider the peculiar 
dependance upon the angular variables, the similarity with the data of a BMS
field theory developed in this paper are rather surprising; thus the overall
picture which emerges from our analysis supports
the conjecture that it is possible to construct an S-matrix in an
asymptotically flat space-time with ``in'' and ``out'' states living respectively on
$\Im^+$ and $\Im^-$ with fields carrying BMS labels. Furthermore we believe
that this idea is not antithetical to the philosophy of this paper where we
have focused our attention on developing a quantum BMS field theory living on
$\Im^+$ or on $\Im^-$ and where we have looked for an implementation of the
holographic correspondence through an analysis of the correlation functions. 
On the contrary, the introduction of the null surface approach 
as a tool to compare bulk and boundary data has shown that the angular
dependence is a necessary feature of a BMS field as in 't Hooft approach. 
Nonetheless a direct construction of such a mapping is still a big task far from 
being complete; as pointed in \cite{Arcioni2}, 
one of the main difficulties in this direction lies in the existence of non 
trivial IR sectors which are labeled by pure supertranslation. As suggested
at the end of section 5 and more in detail in section 6, this peculiar behaviour 
is encoded in the boundary data by means of the specific dependence of BMS
fields upon cut functions with a non vanishing pure supranslational term; thus
we believe that this specific characteristic of the boundary data should be one
of the key aspects in the construction of the S-matrix but a
solution of this task is still not available and this topic is left for
future investigations.

\newpage

\appendix
\section{Elements of white noise as an infinite dimensional calculus}
The aim of this appendix is to give some knowledge of infinite dimensional
calculus and white noise analysis in order to keep the paper self-contained. We
will concentrate mainly on definitions and applications that have been useful
throughout the paper and we leave the details of demonstrations and the
development of deeper applications to foundational books \cite{Hida},
\cite{Kuo2}.\\
Let us consider a real separable Hilbert space $\mathcal{H}$ with norm $||,||$
and let us consider an operator $A$ in $\mathcal{H}$ such that there exists an
orthonormal basis $\left\{\zeta_j\right\}$ satisfying the following conditions:
\begin{enumerate}
\item $A\zeta_j=\lambda_j\zeta_j,\;\;\;\forall j$
\item $1<\lambda_1\leq\lambda_2...\leq\lambda_j\leq...,$
\item there exists $\alpha\in\mathbb{R}_+$ such that
$\sum\lambda_j^{-\alpha}<\infty$.
\end{enumerate}
If, as an example, we consider the Hilbert space $L^2(S^2)$ and the canonical
base of spherical harmonics any operator $J^2+\epsilon$, where $\epsilon>1$ and
where $J$ is the angular momentum, plays the role of $A$.

The white noise space associated to $(\mathcal{H},A)$ can be constructed in the
following way: for each positive integer $p$ define the element
$|\zeta|_p=||A^p\zeta||$ and the set
$\mathcal{E}_p=\left\{\zeta\in\mathcal{H};|\zeta|_p<\infty\right\}$. Let us call
the projective limit of $\left\{\mathcal{E}_p\;\;p>0\right\}$ as $\mathcal{E}$ and 
$\mathcal{E}^\prime$ its dual. This allows us to define the Gelfand triplet
$$\mathcal{E}\subset\mathcal{E}_p\subset\mathcal{H}\subset\mathcal{E}^\prime_p\subset\mathcal{E}^\prime.\;\;\;
p\geq 0$$
Through the Minlos theorem \cite{Kuo} we can introduce the unique probability
measure $\mu\in\mathcal{E}^\prime$ and the complex Hilbert space of
square-integrable functions respect to this measure i.e. $\mathcal{H}^\prime=L^2(\mathcal{H},\mu)$.
From now on this will be our reference space and we will introduce operators
acting on $\mathcal{H}^\prime$. The first step is to define a Gelfand triplet
also for this Hilbert space; this task can be achieved through the Ito-Wiener
lemma which allows us to uniquely express each element
$\phi\in\mathcal{H}^\prime$ as
$$\phi=\sum\limits_{n=0}^\infty I_n(f_n),\;\;\;f_n\in\mathcal{H}_c^{\otimes n}$$
where $\mathcal{H}_c^{\otimes n}$ is the complexification of the symmetric
tensor product of $\mathcal{H}$ and where $I_n$ is the multiple Wiener-integral
(see chapter 3 in \cite{Kuo2} for an analysis). Let us now introduce the
operator $\Gamma(A)$ such that 
$$\Gamma(A)\phi=\sum\limits_{n=0}^\infty I_n(A^{\otimes n}f_n).$$
We can now repeat the construction of a Gelfand triplet defining for each $p>0$
the norm
$$||\phi||_p=||\Gamma(A)^p\phi||,$$
and, consequently,
$$\mathcal{T}_p=\left\{\phi\in
L^2(\mathcal{H})\;\;\;||\phi||_p<\infty\right\}.$$
Introducing the projective limit of $\left\{\mathcal{T}_p\;\;p>0\right\}$ as
$\mathcal{T}$ and its dual space $\mathcal{T}^*$, we have the Gelfand triplet
associated to $\mathcal{H}^\prime$
$$\mathcal{T}\subset\mathcal{T}_p\subset\mathcal{H}^\prime\subset\mathcal{T}^*_p\subset\mathcal{T}^*.$$
Let us now introduce some suitable operations on the above varieties. In
particular let us define the so-called {\it S-transform} of an element
$\Phi\in\mathcal{T}^*$ as the function
\begin{equation}\label{esse}
S\Phi(x)=\langle\langle\Phi,:e^{<\cdot,x>}:\rangle\rangle,
\end{equation}
where $\langle\langle,\rangle\rangle$ is the internal product on
$\mathcal{H}^\prime=L^2(\mathcal{H},d\mu)$ (in order to avoid confusions with the norm $||,||$ in
$\mathcal{H}$) and where 
$$:e^{<x,h>}:=e^{<x,h>-\frac{1}{2}||h||^2}.$$
A second transformation we can introduce is the {\it $\mathcal{T}$-transform}
which assigns to each function $\Phi\in\mathcal{T}^*$ the function
\begin{equation}\label{tau}
\mathcal{T}\Phi(x)=\langle\langle\Phi,e^{i<\cdot,x>}\rangle\rangle.
\end{equation}
Let us now introduce continuous linear operators acting on the space of
test-functions; in particular for any
$\phi\in\mathcal{T}$ we can define the {\it Gateaux derivative} or the
directional derivative along the direction $y\in\mathcal{E}^\prime$ as
\begin{equation}
D_y\phi(x)=\lim_{\epsilon\to 0}\frac{\phi(x+\epsilon y)-\phi(x)}{\epsilon}.
\end{equation}
Let us stress that, given $\eta\in\mathcal{E}$, the operator $D_\eta$ admits a unique
extension by continuity to a continuous linear operator $\tilde{D}_\eta$ on
$\mathcal{T}^*$. \\
A second fundamental operator which has to be defined on $\mathcal{T}$ is the
{\it multiplication operator} $Q_\eta$ ($\eta\in\mathcal{E}$):
\begin{equation}
Q_\eta\phi(x)=<x,\eta>\phi(x).
\end{equation} 
As for the Gateaux derivative, the multiplication operator admits a unique
extension by continuity to a continuous linear operator $\tilde{Q}_\eta$ on
$\mathcal{T}^*$.
A further important concept for the theory of operators acting on Hilbert spaces
is the adjunction i.e., for any operator $J\in\mathcal{T}$, we define the adjoint
operator $J^*\in\mathcal{T}^*$ as
\begin{equation}
\langle\langle
J^*\Phi,\phi\rangle\rangle=\langle\langle\Phi,J\phi\rangle\rangle.\;\;\;\Phi\in\mathcal{T}^*,\phi\in\mathcal{T}
\end{equation}
Moreover, if $J$ is a continuous linear operator, then this property holds also
for the adjoint operator. 

We stress to the reader the difference between the operators $D^*_\eta$ and
$\tilde{D}_y$ albeit both are defined on $\mathcal{T}^*$. This difference can be
further emphasised by the following commutation relation\footnote{Here we only
refer to the non vanishing commutation relations}:
\begin{eqnarray}
\left[\tilde{D}_\eta,D^*_y\right]=<y,\eta>I\;\;{\textrm
on}\;\mathcal{T}^*\;\;\forall\eta\in\mathcal{E},y\in\mathcal{E}^\prime\\
\left[D_y,D^*_\eta\right]=<y,\eta>I\;\;{\textrm on}\;\mathcal{T}\;\;\forall\eta\in\mathcal{E},y\in\mathcal{E}^\prime
\end{eqnarray}
A useful set of identities that holds and relates the multiplication and the
derivative operators is the following:
\begin{gather}
Q_\eta=D_\eta+D^*_\eta,\label{stokes}\\
\tilde{Q}_\eta=\tilde{D}_\eta+D^*_\eta.
\end{gather}
An important consequence of (\ref{stokes}) is the definition of integration by
parts within the realm of Gaussian measures i.e., for any
$\phi,\psi\in\mathcal{T}$,
$$\int\limits_{\mathcal{E}^\prime}\left(D_\eta\phi\right)\psi(x)d\mu(x)=-\int\limits_{\mathcal{E}^\prime}
\phi(x)\left[\left(D_\eta\psi\right)(x)-<x,\eta>\psi(x)\right]d\mu(x).$$
We are now in position to introduce the concept of {\it Fourier transform}
$\hat{\Phi}$ of a generalised function $\Phi\in\mathcal{T}^*$ as the unique
element whose S-transform is 
$$S\hat{\Phi}(x)=(-ix)(S\Phi)e^{-\frac{1}{2}<x,x>},\;\;\;\forall
x\in\mathcal{E}_c$$
where the subscript stands for the complexification of $\mathcal{E}$. At a level
of transformations we can relate the Fourier transform $\mathcal{F}$ to the
other transforms as
\begin{equation}\label{four}
\mathcal{F}=\mathcal{T}^{-1}S.
\end{equation} 
Although the above definition is highly non trivial, $\mathcal{F}$ shares several
common properties with its finite-dimensional counterpart. In particular, when
it is applied to the differential and multiplication operators, the following
identities hold:
\begin{gather}
\mathcal{F}\tilde{D}_\eta=i\tilde{Q}_\eta\mathcal{F},\\
\mathcal{F}D^*_x=-iD^*_x\mathcal{F},\\
\mathcal{F}\tilde{Q}_\eta=i\tilde{D}_\eta\mathcal{F}.
\end{gather}
A last important remark is related to a reiterate application of the Fourier
transform; in order to correctly address this issue, let us introduce the
{\it reflection map} $\rho:\mathcal{T}^*\to\mathcal{T}^*$ such that (see
\cite{Hida})
$$S(\rho\Phi)(x)=S\Phi(-x).$$
Thus the following identity holds:
$$\mathcal{F}=\rho S^{-1}\mathcal{T},$$
and, from (\ref{four}),
$$\mathcal{F}^2=\rho\;\;\; and\;\;\; \mathcal{F}^4=Id.$$
Let us stress that, with the due exception of this appendix, throughout the
paper the domain of definition of all the functions is implicit and it is never
explicitly stated.
\section{Null surface formulation of general relativity}
In this appendix we review the main developments and techniques that
lead to an equivalent formulation of general relativity through characteristic
surfaces since we believe that this approach to the description of the dynamic
of the gravitational field plays a fundamental role in the quest to find an
holographic description in asymptotically flat space-times.

Let us consider an asymptotically flat space-time $M$ with boundary $\Im^\pm$ and a
given metric $g_{ab}$; if we consider a point $x_a\in  M$, we can construct the
light cone with apex $x_a$ through the null geodesic equation
$L(x_a,x^\prime_a)=0$ where $x^\prime_a$ is a generic point on the cone. If we
send $x^\prime_a$ to the boundary $\Im^+$, we can introduce a Bondi coordinate
frame $(u,r=0,z,\bar{z})$ and we can rewrite the light cone equation as
\cite{Kozameh}:
\begin{equation}\label{lcone}
L(x_a,u,z,\bar{z})=0,
\end{equation}
which can be inverted in the variable $u$ as
\begin{equation}\label{fund}
u=Z(x_a,z,\bar{z}),
\end{equation}
which will be the fundamental equation in the null surface description of
general relativity. It is imperative to stress that, since (\ref{fund}) is the
inverse of (\ref{lcone}), a priori it is not globally differentiable or single
valued. Nonetheless, at least locally (i.e. in a suitable open set of
$x^\prime_a=(u,z,\bar{z})$), the function $Z$ is either differentiable either
single valued. Moreover we can draw the following conclusions on the
$Z$-function:
\begin{itemize}
\item keeping fixed the point $x_a$ in (\ref{fund}), this equation describes the
so called {\it light-cone cut} i.e. the intersection $\mathcal{C}_{x_a}$ between
the light cone $N_{x_a}$ with apex $x_a$ and the boundary $\Im^+$ or, more
analytically, $Z$ is a map from the $S^2$-set of light like directions
starting from $x_a$ into a suitable two dimensional surface embedded in $\Im^+$.
\item keeping fixed instead the boundary point $x^\prime_a=(u,z,\bar{z})$, (\ref{fund})
describes the past light-cone of $x^\prime_a$.
\item the function $Z(x_a,z,\bar{z})$ is highly non local.
\item at a geometric level, $Z(x_a,z,\bar{z})$ is a section of the null bundle
$(M^4\times S^2,M^4,\pi)$ where $\pi$ is the trivial projection and where
$$g^{ab}Z_{,a}Z_{,b}=0,$$
i.e. $Z$ is a characteristic surface.
\end{itemize}
Let us stress that the light cone cuts are not regular surfaces except in some
peculiar case as Minkowski space-time. In general, these surfaces can develop
caustics or self-interactions but nonetheless $\mathcal{C}_{x_a}$ will be
always homotopically equivalent to $S^2$ \cite{Kozameh2}. In the language of
fiber bundles this is equivalent to state that the function $Z$ is not a global
section but it is only local and that the light cone cuts can be described as
suitable projections of smooth two-dimensional surfaces embedded in a larger
space i.e. the cotangent bundle $T^*(\Im^+)$.
\par

Since the key idea is to consider (\ref{fund}) as the fundamental variable, a
natural question which arises is if it is possible to construct a metric
$g_{ab}(x)$ (which is the fundamental field in general relativity) starting from
an arbitrary function $Z(x_a,z,\bar{z})$ which also implies that the cuts $\mathcal{C}_{x_a}$
belong only to a certain class of non local function capable of producing a
local field in the bulk $M^4$. The question has been dealt with and it has been 
answered in \cite{Kozameh} where the authors impose suitable conditions both on
the metric components and on the
set of functions $Z(x_a,z,\bar{z})$ in order to correctly associate to each $Z$ 
a non vanishing symmetric tensor field $q_{ab}(x)$ such that
$$q^{ab}(x)Z_{,a}Z_{,b}=0.$$ 
Without entering the details of the derivation, let us only review the main results; 
let us introduce the \emph{edth} derivative operator $\edth$ (see \cite{edth}
for a definition) and the tetrad 
basis $\Theta^i_a=(Z_{,a},\edth Z_{,a},\bar{\edth}Z_{,a},\edth\bar{\edth}Z_{,a})$
in such a way that the metric components can be written as
$$q^{ij}=q^{ab}\Theta_a^i\Theta^j_b.$$
Applying recursively the derivatives $\edth$ and $\bar{\edth}$, we end up
with the following constraints on $q^{ij}$:
\begin{gather}\label{vincmetric}
q^{00}=q^{0+}=q^{0-}=0,\\
\frac{q^{+-}}{q^{01}}=-1\;\;\;\frac{q^{++}}{q^{01}}=-\Lambda_1\;\;\frac{q^{--}}{q^{01}}=-\bar{\Lambda}_1,\\
\frac{q^{+1}}{q^{01}}=-\frac{1}{2}(\bar{\edth}\Lambda_1+\Lambda_1\bar{W})\;\;\;
\frac{q^{-1}}{q^{01}}=-\frac{1}{2}(\edth\bar{\Lambda}_1+\bar{\Lambda}_1 W,),\\
\edth\ln q^{01}=W\;\;\;\bar{\edth}\ln q^{01}=\bar{W},\\
\frac{q^{11}}{q^{01}}=-2\epsilon(\Lambda_0,\Lambda_1,\Lambda_-),
\end{gather}
where $\epsilon$ is a real function (see appendix A in \cite{Kozameh} for an
explicit expression), $\Lambda(x,z,\bar{z})=\edth^2 Z$, $\Lambda_{,a}=\Lambda_i\Theta^i_a$
and $W=\frac{\partial\Lambda_1-2\Lambda_-}{\Lambda_1}+\partial\ln P$ with
$P=(1-\Lambda_1\bar{\Lambda}_1)^{-1}$.
The cut function $Z(x_a,z,\bar{z})$ or, more properly, its second derivative has
to satisfy the following equations:
\begin{gather}
\Lambda_+=W-\frac{1}{2}\left(\Lambda_1\bar{W}+\bar{\edth}\Lambda_1+\edth\ln
P\right)\;\;\;{\textrm and }\;{\textrm c.c.},\\
\edth\epsilon+\epsilon(W-2\Pi_1)+(1-\frac{\Pi_+}{2})(\edth\Lambda_1+\Lambda_1\bar{W})+\notag\\
-\frac{1}{2}(\edth\bar{\Lambda}_1+\bar{\Lambda}_1W)\Pi_-+\Pi_0=0\;\;\;{\textrm and }\;{\textrm c.c.},
\end{gather}
where $\Pi$ is a function of $Z$ and its derivatives (still see appendix A in
\cite{Kozameh} for a definition). Moreover if we want to define a Lorentzian 
metric, we have to impose a further condition:
\begin{equation}
\det\left(q_{ab}\right)<0\Longrightarrow\det\left(q_{ij}\right)=(q^{01})^4P>0,
\end{equation}
or equivalently $P>0\Longleftrightarrow |\Lambda_1|<1$.\\
Thus the set of functions satisfying the above conditions allow us to write a
Lorentzian metric and, consequently, the connection and curvature tensor as 
functionals only of the cut equation. Nonetheless this
condition is not sufficient since this does not grant us that the metric
satisfies Einstein equations. In order to impose also this requirement, one has to
realize \cite{Frittelli2}, \cite{Frittelli5} that, in order to reformulate general relativity in
terms of equations for families of surfaces, one also needs to introduce a second
function $\Omega(x^a,z,\bar{z})$ that acts as a conformal factor turning the
conformal metrics constructed from the characteristic surfaces
$u=Z(x^a,z,\bar{z})$ into an Einstein metric. Thus the vacuum Einstein equations
become (see as an example \cite{Frittelli3})
\begin{equation}\label{eins}
D^2\Omega=Q(\Lambda)\Omega,
\end{equation}  
where $Q=-\frac{1}{4q}(D\bar{\Lambda}_1
D\Lambda_1)-\frac{3}{8q^2}(Dq)^2+\frac{1}{4q}D^2q$,
$q=1-\Lambda_1\bar{\Lambda}_1$ and
$D=\frac{\partial}{\partial(\edth\bar{\edth}Z)}$. Moreover both $\Lambda$ and $\Omega$ have to satisfy the following consistency equations:
\begin{gather}
\edth\Omega= W(\Lambda_{,1})\Omega,\label{uno}\\
\edth\Lambda_{,1}-2\Lambda_{,-}=(W+\edth(\ln q))\Lambda_{,1},\label{due}\\
\edth^2\Lambda=\bar{\edth}^2\Lambda\label{tre}.
\end{gather}
At this stage we need to emphasise that, although the above set of differential 
equations is rather difficult to solve in a general scenario, it can be
simplified for regular asymptotically flat space-times. In particular, as shown
in details in \cite{Frittelli}, if we introduce the asymptotic Bondi shear
$\sigma(u,z,\bar{z})$, (\ref{uno}), (\ref{due}), (\ref{tre}) can be
substituted by the equation:
\begin{equation}
\edth^2\bar{\edth}^2Z=\bar{\edth}^2\sigma_R+\edth^2\sigma_R+N[Z,\Omega],
\end{equation}  
where $N$ is a complicated function of $\Lambda$ and its derivatives
(see \cite{Frittelli} for the explicit expression) and $\sigma_R$ represents
the freely chosen Bondi shear with the $u-$variable substituted by
$Z(x_a,z,\bar{z})$.

\bigskip

\begin{center}
\noindent\textbf{Acknowledgements}
\end{center}

\vspace{0.1cm}
\noindent This work was supported by a research grant from the Department of Nuclear and
Theoretical physics - University of Pavia.
The author is grateful to M. Carfora, O. Maj and, in particular, to G. Arcioni
for useful discussions and comments.


\thebibliography{100}
\bibitem{'tHooft}
G.~'t Hooft,
\emph{``Dimensional reduction in quantum gravity,''}
arXiv:gr-qc/9310026.

\bibitem{Aharony}
O.~Aharony, S.~S.~Gubser, J.~M.~Maldacena, H.~Ooguri and Y.~Oz,
\emph{``Large N field theories, string theory and gravity,''}
Phys.\ Rept.\  {\bf 323} (2000) 183
[arXiv:hep-th/9905111].

\bibitem{deBoer}
J.~de Boer, L.~Maoz and A.~Naqvi,
\emph{``Some aspects of the AdS/CFT correspondence,''}
arXiv:hep-th/0407212.

\bibitem{deBoer2}
J.~de Boer and S.~N.~Solodukhin,
\emph{``A holographic reduction of Minkowski space-time,''}
Nucl.\ Phys.\ B {\bf 665} (2003) 545
[arXiv:hep-th/0303006].

\bibitem{Solodukhin}
S.~N.~Solodukhin,
\emph{``Reconstructing Minkowski space-time,''}
arXiv:hep-th/0405252.

\bibitem{Alvarez}
E.~Alvarez, J.~Conde and L.~Hernandez,
\emph{``Goursat's problem and the holographic principle,''}
Nucl.\ Phys.\ B {\bf 689} (2004) 257
[arXiv:hep-th/0401220].

\bibitem{Arcioni}
G.~Arcioni and C.~Dappiaggi,
\emph{``Exploring the holographic principle in asymptotically flat
space-times
via the BMS group,''}
Nucl.\ Phys.\ B {\bf 674} (2003) 553
[arXiv:hep-th/0306142].

\bibitem{Arcioni2}
G.~Arcioni and C.~Dappiaggi,
\emph{``Holography in asymptotically flat space-times and the BMS group,''}
Class.\ Quant.\ Grav.\  {\bf 21} (2004) 5655
[arXiv:hep-th/0312186].

\bibitem{Sternberg}
S.~Sternberg,
\emph{``Minimal Coupling And The Symplectic Mechanics Of A Classical Particle In The
Presence Of A Yang-Mills Field,''}
Proc.\ Nat.\ Acad.\ Sci.\  {\bf 74} (1977) 5253.

\bibitem{Guillemin}
V.~Guillemin and S.~Sternberg,
\emph{``On The Equations Of Motion Of A Classical Particle In A Yang-Mills Field And
The Principle Of General Covariance,''}
Hadronic J.\  {\bf 1} (1978) 1.

\bibitem{Weinstein}
A.~Weinstein,
\emph{``A Universal Phase Space For Particles In Yang-Mills Field,''}
Lett.\ Math.\ Phys.\  {\bf 2} (1978) 417.

\bibitem{Wald2} R.~Wald \emph{``General Realtivity''} (1984) University of Chicago Press

\bibitem{Geroch} R.~Geroch \emph{``Asymptotic structures of spacetime} ed. P. Esposito and L. Witten (New York: Plenum) 1977

\bibitem{Geroch2} R.~Geroch, B.~Xanthopoulos \emph{``Asymptotic simplicity is
stable''} J. Math. Phys {\bf 19} (1978) 714.

\bibitem{Mc1} P.J. McCarthy: \emph{''Representations of the
Bondi-Metzner-Sachs
group I''} Proc. R. Soc. London {\bf A330} 1972 (517).


\bibitem{Mc4} P.J. McCarthy: \emph{``The Bondi-Metzner-Sachs in the nuclear
topology''} Proc. R. Soc. London {\bf A343} 1975 (489),


\bibitem{Mc2} P.J. McCarthy: \emph{''Representations of the
Bondi-Metzner-Sachs
group II''} Proc. R. Soc. London {\bf A333} 1973 (317),

\bibitem{Mc7} P.J. McCarthy: \emph{``Real and complex symmetries in quantum
gravity, irreducible representations, polygons, polyhedra and the A.D.E.
series''}
Phil. Trans. R. Soc. London A {\bf 338} (1992) 271.

\bibitem{Melas} E. ~Melas, \emph{``The BMS group and generalized gravitational
instantons''} J. Math. Phys. {\bf 45} (2004) 996

\bibitem{Lee}
J.~Lee and R.~M.~Wald,
\emph{``Local Symmetries And Constraints,''}
J.\ Math.\ Phys.\  {\bf 31} (1990) 725.

\bibitem{Chernoff}
P.R. ~Chernoff and J.E. ~Marsden,
\emph{``Properties of Infinite Dimensional Hamiltonian Systems,''}
Springer-Verlag (1974),

\bibitem{streubel}
A.~Ashtekar and M.~Streubel,
\emph{``Symplectic Geometry Of Radiative Modes And Conserved Quantities At
Null
Infinity,''}
Proc.\ Roy.\ Soc.\ Lond.\ A {\bf 376} (1981) 585.

\bibitem{Hida} T. ~Hida, H. ~Kuo, J. ~Potthoff, L. ~Streit, \emph{``White noise:
An infinite dimensional calculus''} Kluwer Academic Publisher (1993)

\bibitem{Kuo2} Hui-Hsiung ~Kuo, \emph{``White noise distribution theory''} CRC
Press (1996)

\bibitem{Barut} A.O. Barut, R. Raczka: \emph{``Theory of group representation and
applications''} World Scientific 2ed (1986),

\bibitem{Schmid} R. ~Schmid, A. ~Simoni, \emph{``On infinite-dimensional
variational principles with constraints''} J. Math. Phys. {\bf 30} (1989) 1171

\bibitem{Kuo} Hui-Hsiung ~Kuo, \emph{``Gaussian measures in Banach spaces''}
Springer-Verlag (1975),

\bibitem{Peskin} M. E. ~Peskin, D. V. ~Schroeder: \emph{``An introduction to
quantum field theory''} Perseus Books (1995).

\bibitem{Asorey}
M.~Asorey, L.~J.~Boya and J.~F.~Carinena,
\emph{``Covariant Representations In A Fiber Bundle Framework,''}
Rept.\ Math.\ Phys.\  {\bf 21} (1986) 391.

\bibitem{asymptoticquantization} A.~Ashtekar,
\emph{``Asymptotic Quantization: Based On 1984 Naples Lectures,''} 
Bibliopolis, Naples (1987).

\bibitem{workinprogress}
G. ~Arcioni, C. ~Dappiaggi,
\emph{``work in progress''}.

\bibitem{Kozameh}C. ~Kozameh, E. ~T. ~Newman, \emph{``Theory of light cone
cuts of null infinity''} J. Math. Phys. {\bf 24}
(1983) 2481,

\bibitem{Frittelli6}
S.~Frittelli, E.~T.~Newman and G.~Silva-Ortigoza,
\emph{``The eikonal equation in flat space: Null surfaces and their  singularities.
I,''}
J.\ Math.\ Phys.\  {\bf 40} (1999) 383
[arXiv:gr-qc/9809019].

\bibitem{Frittelli7}
S.~Frittelli, E.~T.~Newman and G.~Silva-Ortigoza,
\emph{``The eikonal equation in asymptotically flat space-times,''}
J.\ Math.\ Phys.\  {\bf 40} (1999) 1041.

\bibitem{Frittelli8}
S.~Frittelli, E.~T.~Newman
\emph{``Poincar\'e pseudosymmetries in asymptotically flat spacetimes''} in
\emph{On Einstein's path} Springer, New York (1999) 227

\bibitem{Frittelli}S. ~Frittelli, C. ~Kozameh, E. ~T. ~Newman, \emph{``Dynamics
of light cone cuts of null infinity''} Phys. Rev. D {\bf 56} (1997) 4729,

\bibitem{Frittelli4}S. ~Frittelli, E. ~T. ~Newman, \emph{``
Pseudo-Minkowskian coordinates in asymptotically flat space-times''} 
Phys. Rev. D {\bf 55} (1997) 1971.

\bibitem{Kozameh2} C. ~Kozameh, P. W. ~Lamberti, O. ~Reula: \emph{``Global
aspects of light cone cuts''} J. Math. Phys. {\bf 32} (1991) 3423.

\bibitem{Joshi}P. S. ~Joshi, C. ~Kozameh, E. ~T. ~Newman, \emph{``Light cone
cuts of null infinity in Schwarzschild geometry''} J. Math. Phys. {\bf 24}
(1983) 2490, 

\bibitem{Joshi2}P. S. ~Joshi, C. ~Kozameh, E. ~T. ~Newman, 
\emph{``Light-cone cuts of $I\sp +$ for charged Kerr geometry''} 
Gen. Rel. Grav. {\bf 16} (1984) 1157,

\bibitem{edth}R.~Penrose and W.~Rindler,
\emph{``Spinors And Space-Time. Vol. 2: Spinor And Twistor Methods In Space-Time
Geometry,''} Cambridge University Press (1987)

\bibitem{Frittelli2}S. ~Frittelli, C. ~Kozameh, E. ~T. ~Newman, \emph{``
GR via characteristic surfaces''} J. Math. Phys. {\bf 36} (1995) 4904,

\bibitem{Frittelli5}
S.~Frittelli, C.~Kozameh and E.~T.~Newman,
\emph{``Lorentzian metrics from characteristic surfaces,''}
J.\ Math.\ Phys.\  {\bf 36} (1995) 4975
[arXiv:gr-qc/9502025].

\bibitem{Frittelli3}S. ~Frittelli, C. ~Kozameh, E. ~T. ~Newman, \emph{``
On the dynamics of characteristic surfaces''} J. Math. Phys. {\bf 36} (1995)
6397,

\bibitem{'tHooft2}
G.~'t Hooft,
\emph{``The scattering matrix approach for the quantum black hole: An
overview,''}
Int.\ J.\ Mod.\ Phys.\ A {\bf 11} (1996) 4623
[arXiv:gr-qc/9607022].

\bibitem{Arcioni4}
G.~Arcioni,
\emph{``On 't Hooft's S-matrix ansatz for quantum black holes,''}
JHEP10(2004)032,
[arXiv:hep-th/0408005].
\end{document}